\documentclass[reqno]{amsart}

\usepackage{booktabs}
\usepackage{IEEEtrantools}
\usepackage{latexsym}
\usepackage{graphicx}       
\usepackage{mathrsfs}
\usepackage{hyperref}
\usepackage{dsfont}

\def\baselinestretch{1}

\usepackage[noadjust]{cite}
\usepackage{extarrows}

\usepackage{paralist}
\usepackage{capt-of}
\usepackage{xcolor}
\usepackage{diagbox}
\usepackage{cite}
\usepackage{subfigure}
\usepackage{amsmath,amssymb,amsfonts}
\usepackage{textcomp}
\usepackage{wasysym}
\usepackage{caption}

\usepackage{setspace}
\usepackage{cite}
\usepackage{color}

\usepackage{comment}

\usepackage{diagrams}
\usepackage{color}
\usepackage{cite}
\usepackage{subfigure}
\usepackage{amssymb,amsfonts,amsmath}
\renewcommand{\theequation}{\arabic{equation}}
\usepackage{graphicx,xcolor}
\usepackage{subcaption}
\usepackage{stfloats}
\usepackage{enumitem}
\usepackage{flushend}
\usepackage{cite}
\usepackage{amsmath,amssymb,amsfonts}
\usepackage{graphicx}
\usepackage{textcomp}
\usepackage{wasysym}
\usepackage{caption}
\usepackage{apptools}

\usepackage{color}        
\usepackage{setspace}
\usepackage{microtype}
\usepackage{overpic}
\usepackage{booktabs} 

\usepackage{paralist}
\usepackage{capt-of}
\usepackage{xcolor}
\usepackage[export]{adjustbox}
\usepackage{chngcntr}
\usepackage[normalem]{ulem}

\usepackage{amssymb,latexsym,amsfonts}
\usepackage{paralist}
\usepackage{eso-pic}
\usepackage{epsfig}
\usepackage{mathtools}
\usepackage{epstopdf}
\usepackage{mathtools}
\usepackage{lineno}
\allowdisplaybreaks

\newcommand{\N}{\mathbb{N}}%
\newcommand{\R}{\mathbb{R}}%

\usepackage{tikz}
\usepackage[many]{tcolorbox}
\usetikzlibrary{calc}

\usetikzlibrary{positioning,shapes}
\usetikzlibrary{arrows}

\tcbuselibrary{skins}

\AtAppendix{\counterwithin{theorem}{section}}

\usepackage{subfigure}
\usepackage{amsmath,amsfonts}
\usepackage{graphicx}
\usepackage{textcomp}
\usepackage{wasysym}
\usepackage{caption}
\usepackage{apptools}
\usepackage{multicol,lipsum}

\usepackage{color}        
\usepackage{setspace}
\usepackage{microtype}
\usepackage{overpic}
\usepackage{booktabs} 

\usepackage{paralist}
\usepackage{capt-of}
\usepackage{xcolor}
\usepackage[export]{adjustbox}
\usepackage{chngcntr}
\usepackage[normalem]{ulem}

\usepackage{latexsym,amsfonts}
\usepackage{paralist}
\usepackage{eso-pic}
\usepackage{epsfig}
\usepackage{mathtools}
\usepackage{epstopdf}
\usepackage{mathtools}
\usepackage{lineno}
\usepackage[linesnumbered,ruled,algo2e]{algorithm2e}

\usepackage{tikz}
\usepackage[many]{tcolorbox}
\usetikzlibrary{calc}
\usetikzlibrary{backgrounds,fit}
\usetikzlibrary{shapes.arrows}

\usepackage{url}
\usepackage{pgf}
\usetikzlibrary{positioning, arrows.meta,automata}

\usetikzlibrary{positioning,shapes}
\usetikzlibrary{arrows}

\tcbuselibrary{skins}
\newtcolorbox{resp}[1][]{%
enhanced jigsaw,%
colback=gray!5!white,%
colframe=gray!80!black,%
size=small,%
boxrule=1pt,%
halign title=flush center,%
coltitle=black,%
breakable,%
drop shadow=black!50!white,%
attach boxed title to top left={xshift=1cm,yshift=-\tcboxedtitleheight/2,yshifttext=-\tcboxedtitleheight/2},%
minipage boxed title=3cm,%
boxed title style={%
	colback=white,%
	size=fbox,%
	boxrule=1pt,%
	boxsep=2pt,%
	underlay={%
		\coordinate (dotA) at ($(interior.west) + (-0.5pt,0)$);
		\coordinate (dotB) at ($(interior.east) + (0.5pt,0)$);
		\begin{scope}[gray!80!black]
			\fill (dotA) circle (2pt);
			\fill (dotB) circle (2pt);
		\end{scope}
	}%
},%
#1%
}

\definecolor{myco}{rgb}{0.55, 0.0, 0.63}

\AtAppendix{\counterwithin{theorem}{section}}

\newtheorem{theorem}{Theorem}[section]
\newtheorem{lemma}[theorem]{Lemma}
\newtheorem{problem}[theorem]{Problem}

\newtheorem{corollary}[theorem]{Corollary}

\newtheorem{definition}[theorem]{Definition}
\newtheorem{example}{Example}

\newtheorem{assumption}[theorem]{Assumption}
\numberwithin{equation}{section}   

\begin{document}
	
\begin{abstract}
Diagnosability is a system theoretical property characterizing whether fault occurrences in a system can always be detected within a finite time. In this paper, we investigate the verification of diagnosability for cyber-physical systems with continuous state sets.
We develop an abstraction-free and automata-based framework to verify (the lack of) diagnosability, leveraging a notion of \emph{hybrid barrier certificates}. 
To this end,  we first construct a ($\delta$,$K$)-deterministic finite automaton that captures the occurrence of faults targeted for diagnosis. Then, the verification of diagnosability property is converted into a safety verification problem over a product system between the automaton and the augmented version of the dynamical system.  We demonstrate that this verification problem can be addressed by computing hybrid barrier certificates for the product system. To this end, we introduce two systematic methods, leveraging sum-of-squares programming and counter-example guided inductive synthesis to search for such certificates. Additionally, if the system is found to be diagnosable, we propose methodologies to construct a diagnoser to identify fault occurrences online. Finally, we showcase the effectiveness of our methods through a case study. 
\end{abstract}

\title[Verification of Diagnosability for Cyber-Physical Systems]{Verification of Diagnosability for Cyber-Physical Systems: A Hybrid Barrier Certificate Approach}

\author{Bingzhuo Zhong$^{1*}$}
\author{Weijie Dong$^{2*}$}
\author{Xiang Yin$^{2}$}
\author{Majid Zamani$^{3}$}

\thanks{$^*$Both authors contributed equally to this research.}

\address{$^1$Thrust of Artificial Intelligence, Information Hub, Hong Kong University of Science and Technology (Guangzhou), China}
\email{bingzhuoz@hkust-gz.edu.cn}

\address{$^2$ Department of Automation, Shanghai Jiao Tong University, China.}
\email{\{wjd\_dollar,yinxiang\}@sjtu.edu.cn}

\address{$^3$Department of Computer Science, University of Colorado Boulder, USA}
\email{majid.zamani@colorado.edu}
\maketitle

\section{Introduction}
\subsection{Motivation}
Complex cyber-physical systems (CPSs), including intelligent transportation systems and manufacturing systems, are fundamental to safety-critical infrastructure. 
Nevertheless, these systems often experience \emph{failures} that are challenging to detect due to the multitude of internal components and subsystems operating concurrently. 
This complexity renders \emph{fault diagnosis} and \emph{detection} essential yet challenging tasks, which are crucial for ensuring safety and functionality over complex CPSs.

To tackle these challenges, assessing the \emph{diagnosability} of complex CPSs is pivotal. 
Diagnosability, which was first introduced in \cite{sampath1995diagnosability}, 
is a type of information-flow properties that determines whether or not faults within a system can be conclusively identified within a finite time frame after their occurrences. 
Analyzing diagnosability provides several advantages~\cite{lu2007system}. 
For example, in the design phase of CPSs, results from diagnosability analysis help optimizing sensor placement and the development of built-in tests, enhancing later fault diagnosis and detection. 
For operational systems, these results inform improvements in diagnostics and maintenance scheduling. 
Therefore, a deeper understanding on the diagnosability of CPSs allows for a cost-effective assessment of trade-offs among reliability, functionality, and diagnosability.

Building upon the foundational study in \cite{sampath1995diagnosability}, the field of diagnosability analysis has been extensively investigated, especially within the context of discrete event systems (DES), which are characterized by event-driven dynamics and a discrete set of states, including 
Petri nets \cite{basile2017diagnosability,lefebvre2007diagnosis,hu2021diagnosability,ma2021marking,pencole2022diagnosability,ran2018codiagnosability,yin2017decidability},
stochastic automata \cite{thorsley2005diagnosability,chen2023probabilistic,yin2019robust}, and
finite state automata \cite{lin2017n,ma2023verification,takai2017generalized}. 
Note that these works mainly focus on exploring diagnosability utilizing event-based observation models, whose outputs are symbolically distinct and tagged with unique labels. 
However, in many real-world applications, system outputs are typically continuous-space physical signals rather than discrete event sequences. 
Due to the interference of measurement noise in these signals, identifying two continuous signals with absolute accuracy are difficult, making it challenging to apply the traditional framework of diagnosability devised for DES to systems with continuous-space outputs. 

To bridge this gap, research efforts have expanded the concept of diagnosability to encompass systems operating in continuous state sets \cite{pola2017approximate,di2011verification,deng2016verification,bayoudh2008hybrid}.
Specifically, \cite{bayoudh2008hybrid} explored diagnosability in hybrid systems that exhibit both continuous and discrete dynamics by representing mode switches as discrete events. 
The authors in \cite{di2011verification} introduced and verified a new notion of diagnosability for hybrid systems by abstracting them into timed automata. 
This method was further refined in \cite{deng2016verification}, which incorporated assessments of the impacts of time delays and measurement uncertainties on fault detection. 
Additionally, \cite{pola2017approximate} proposed the concept of \emph{approximate diagnosability} for metric systems, accounting for inprecision in output measurements. 
The assessment of this approximate diagnosability was then conducted using a finite-state model, established through an approximate simulation relation \cite{pola2017approximate}. 
However, this method faces significant challenges due to the curse of dimensionality, particularly because it requires discretizing the continuous state and input sets to build a finite abstraction of the system.

\subsection{Contribution} 
In this paper, we explore the concept of \emph{$\delta$-approximate $K$-diagnosability} in the context of discrete-time control systems, extending the notion of $K$-diagnosability, as introduced in \cite{dallal2013most}, to metric systems. 
This requires that all faults should be detectable within a maximum of $K$ time steps after they occur. 
As we do not have any prior knowledge of whether or when faults will occur, the primary challenge in abstraction-free verification lies in accurately tracking faults and recording the time elapsed since their occurrence without relying on a finite abstraction of the original system.

To address these challenges, we construct a deterministic finite automaton (\emph{$(\delta,K)$-DFA}) that effectively monitors fault occurrences and the elapsed time since their occurrence, independent of when the fault occurs. 
Subsequently, through integrating this automaton with an augmented version of the discrete-time control system to construct a product system, we establish the necessary and sufficient conditions under which the control system achieves $\delta$-approximate $K$-diagnosability.
Note that these conditions also provide a uniform framework for verifying (the lack of) $\delta$-approximate $K$-diagnosability over both discrete-event systems and continuous-space systems, regardless of using discretization-based or discretization-free approaches.
Based on these conditions, the verification of (the lack of) $\delta$-approximate $K$-diagnosability is refined into a safety verification problem over the product system, which is essentially a hybrid system.
To solve such a safety verification problem, we leverage the notion of \emph{hybrid barrier certificates} introduced in~\cite{murali2023co} and propose two methodologies for computing these certificates, grounded in the counterexample-guided inductive synthesis (CEGIS) framework and sum-of-squares (SOS) programming. 
Additionally, we propose the construction of a diagnoser capable of detecting faults when the system is shown to be $\delta$-approximate $K$-diagnosable.

A limited subset of the results in this paper has been presented in~\cite{zhong2024verification}. 
Here, we provide detailed proofs of the results in~\cite{zhong2024verification} that were omitted. 
Apart from the verification of diagnosability, here, we also propose methodologies to verify \emph{the lack of} $\delta$-approximate $K$-diagnosability via \emph{hybrid barrier certificates}.
Accordingly, we propose two computational schemes, namely CEGIS-based and SOS-based schemes, to compute hybrid barrier certificates to verify (the lack of) diagnosability.
Furthermore, construction of a diagnoser that detects fault is provided, which was missing in~\cite{zhong2024verification}. 

\subsection{Related Works}
Since diagnosability is a class of information-flow properties, it is important to highlight various different approximate notions of information-flow properties. For example, to enhance security measures, the concept of \emph{approximate opacity} was proposed for metric systems in \cite{yin2020approximate} and subsequently extended for discrete-time stochastic systems in \cite{liu2020notion}. 
Focusing on state estimation, \emph{approximate observability} was investigated for discrete-time nonlinear systems in \cite{pola2023approximate}. 
These results typically require the construction of finite abstractions from the original systems, which can lead to scalability issues due to the discretization of continuous state and input sets. 
To overcome these limitations, discretization-free methods were introduced in \cite{kalat2021modular,liu2020verification}, which employ barrier certificates to ascertain approximate opacity.  It is crucial to note that these methodologies essentially address different information-flow properties than those discussed in this paper. 

Additionally, we would also like to mention the results in~\cite{zhao2024unified}, which showed that diagnosability over DES can be formulated as HyperLTL properties by proper system transformation over the original DES, and in~\cite{anand2021verification}, which focused on a class of HyperLTL property called \emph{conditional invariance}.
They showed that the verification of such properties can be solved by exploiting the automata-based structures associated with the complements of these properties.
However, $\delta$-approximate $K$-diagnosability does not fall into the class of conditional invariance.
Moreover, it is still an open problem for how to perform the system transformation introduced in~\cite{zhao2024unified} over the DES to systems with continuous or hybrid state and input sets without building finite abstractions. To the best of our knowledge, the current work as well as its preliminary version~\cite{zhong2024verification} are the first to propose methodologies that does not require abstractions for verifying $\delta$-approximate $K$-diagnosability for discrete-time continuous-space systems.

\subsection{Organization}
The remainder of the paper is organized as follows. 
Section \ref{sec:Problem Formulation} introduces notations, foundational models, and problem formulation of this paper. 
Section \ref{sec3} details the construction of a ($\delta$, $K$)-deterministic finite automaton that is employed to verify the $\delta$-approximate $K$-diagnosability property, leveraging the notion of hybrid barrier certificates.
Section \ref{sec4} elaborates two methods for calculating hybrid barrier certificates, specifically through the use of the CEGIS framework and SOS programming techniques. 
In Section \ref{sec:Diagnoser}, we describe the construction of a diagnoser based on the system satisfying the $\delta$-approximate $K$-diagnosability conditions. 
A case study demonstrating the efficacy of our approaches is presented in Section \ref{sec:case}. 
Lastly, the paper concludes with Section \ref{sec:conclusion}.

\section{Problem Formulation}\label{sec:Problem Formulation}

\subsection{Notations}
We denote by $\R$ and $\N$ the set of real numbers and non-negative integers,  respectively. 
These symbols are annotated with subscripts to restrict them in the usual way, e.g., $\R_{>0}$ denotes the set of positive real numbers. 
For $a,b\in\mathbb{R}$ (resp.\ $a,b\in\mathbb{N}$) with $a\leq b$, the closed, open and half-open intervals in $\mathbb{R}$ (resp.\ $\mathbb{N}$) are denoted by $[a,b]$, $(a,b)$, $[a,b)$, and $(a,b]$, respectively. 
Moreover, $\mathbb{R}^{n\times m}$, with $n,m\in \mathbb{N}_{\geq 1}$, denotes the vector space of real matrices with $n$ rows and $m$ columns.
Given $N \in \mathbb N_{\ge 1}$ vectors $x_i \!\in\! \mathbb R^{n_i}$, with $i\!\in\! [1;N]$, $n_i\!\in\! \mathbb N_{\ge 1}$, and $n \!= \!\sum_i n_i$, we denote the concatenated vector in $\mathbb R^{n}$ by $x \!=\! [x_1;\!\ldots\!;x_N]$ and the Euclidean norm of $x$ by $\Vert x\Vert$.
Given a set $X$, we denote by $2^X$ the power set of $X$.
Given a matrix $A$, we denote by 
$A^\top$
and $A_r(i)$  
the transpose 
and the $i$-th row of $A$, respectively.
Considering a set $Y$, $Y^*$ denotes the Cartesian product of a finite number of $Y$.
Given sets $X$ and $Y$, the complement of $X$ with respect to $Y$ is defined as $Y \setminus X \!=\! \{x \!\in\! Y : x \!\notin\! X\}.$
Given functions $f\!:\! X \!\rightarrow \!Y$ and $g\!:\! A \!\rightarrow\! B$, we define $f \times g : X \times A \rightarrow Y \times B$ as $(f \times g)(x,a)=(f(x),g(a))$.

\subsection{Preliminaries}
In this paper, we focus on  discrete-time control systems defined as below.
\begin{definition}\label{def:sys1}
    A \emph{discrete-time control system} (dt-CS) is a tuple $\Sigma:=(X, X_0,X_F, U,f,Y,h)$, where $X\subset \mathbb{R}^n$, $X_0 \subseteq X $, $X_F \subset X$, $U \subset \mathbb{R}^m$, and $Y\subset \mathbb{R}^q$ denote the state set, initial state set, faulty state set, input set, and output set, respectively, which are bounded.  
	Moreover, the function $f:  X\times  U \rightarrow  X$ and $h: X \rightarrow  Y$ are the state transition function and the output function, respectively. 
\end{definition}

Equivalently, a dt-CS $\Sigma$ can be described by 
\begin{align}\label{eq:2}
	\Sigma:\left\{
	\begin{array}{rl}
		x(k+1)&= f(x(k),\nu(k)),\\
		y(k)&=h(x(k)), \quad \quad \quad k\in\mathbb{N},
	\end{array}
	\right.
\end{align}
where $x(k)\in X$, $\nu(k)\in U$, and $y(k)\in Y$ are the state, input, and output of the system at time step $k$, respectively.
Additionally, we denote by $\nu: = (\nu(0),\ldots,\nu(k),\ldots)$ an input run of $\Sigma$, 
by $\mathbf{x}_{x_0,\nu}:= (x(0),\ldots,x(k),\ldots)$ a state run of $\Sigma$ starting from initial state $x_0$ under input run $\nu$, i.e., $x(0)=x_0$, $x(k+1)=f(x(k),\nu(k))$, $\forall k\in \mathbb{N}$, and by $\mathbf{y}_{x_0,\nu}$ the output run corresponding to $\mathbf{x}_{x_0,\nu}$.

A state run $\mathbf{x}_{x_0,\nu}$ is deemed a \emph{faulty} run if it visits the set of faulty states $X_F\subset X$, which means there exists  $i \in \mathbb{N}$ such that $\mathbf{x}_{x_0,\nu}(i) \in X_F$. 
Here, we say that a \emph{fault occurs} if the system's run visits the faulty state set \emph{for the first time}. 
It is important to note that $X_F$ can represent either a physically unsafe region or a region where a task violation is detected. 
In the interest of safety, it becomes imperative to \emph{diagnose} the occurrence of a fault in a timely manner, preventing the possibility of cascading disasters. 
It is worth mentioning that the repair of faults is not considered in this work. 
In other words, once the system enters a faulty state, it remains faulty indefinitely, even if it subsequently exits the faulty state set. 

In most cases, the output run cannot be measured perfectly due to measurement noises. 
Therefore, we assume that the output run can only be observed imprecisely. 
Considering the actual output run $\mathbf{y}_{x_0,\nu}$,
one has 
\begin{align}
    ||\mathbf{y}_{x_0,\nu}(k)-\mathbf{y}_{x_0,\nu}^{\delta}(k)||\leq \delta, \forall k\in \mathbb{N},\label{imprecise}
\end{align}
in which $\delta\in\mathbb{R}_{\geq 0}$ represents the observation precision and $\mathbf{y}_{x_0,\nu}^{\delta}$ is the actual observation associated with $\mathbf{y}_{x_0,\nu}$ that is available.
Given an observation precision $\delta$, 
a fundamental question is whether or not faults in the dt-CS as in Definition~\ref{def:sys1} can always be diagnosed. 
This leads to the notion of \emph{$\delta$-approximate $K$-diagnosability}, which is formally defined below.
\begin{definition}\label{def:diagnosability}
	Consider a dt-CS $\Sigma=(X, X_0,X_F, $  $U,f,Y,h)$. 
	Given a constant $\delta \in\mathbb{R}_{>0}$, and $K \in\mathbb{N}$, $\Sigma$ is said to be \emph{$\delta$-approximate $K$-diagnosable}, if for all $ \mathbf{x}_{x_0,v}:= (x_0,\ldots,x_k,\ldots)$ such that $x_0\in X_0$, $x_k \in X_F$, and $x_i\notin X_F$, $\forall i \in [0,k-1]$; and for all $ \hat{\mathbf{x}}_{\hat{x}_0,\hat{v}}:= (\hat{x}_0,\ldots,\hat{x}_k,\ldots)$ such that $\hat{x}_0\in X_0$, and $\hat{x}_i\notin X_F$, $\forall i\in[0,k+K]$, 
	one has $\max_{i\in [0,k+K]} \vert\vert  h(x_i)- h(\hat{x}_i)\vert \vert > \delta$.  
\end{definition}

Intuitively, $K\in \mathbb{N}$ represents the maximal time window following a fault occurrence at time $k$, within which the fault must be diagnosed.
To verify whether or not a dt-CS is $\delta$-approximate $K$-diagnosable, the notion of \emph{deterministic finite automata} \cite{Baier2008Principles} is leveraged, which is  introduced below.
\begin{definition} \emph{(DFA)}
	A deterministic finite automata (DFA) is a tuple $\mathcal{A}\ =\!(Q, q_0, \Pi,\tau, F)$, in which $Q$ is a finite set of states, $q_0\!\in\! Q$ is the initial state, $\Pi$ is a finite set of alphabet, $\tau : Q\times\Pi \rightarrow Q$ is a transition function, and $F\!\subseteq\! Q$ is a set of accepting states. 
\end{definition}

Considering the transition function $\tau$, we further define a function $\text{Nxt}: Q\!\rightarrow\!  2^Q$ by:
\begin{align}
   \text{Nxt}(q)\! :=\!\{ q'\in Q: \exists \sigma \!\in\!\Pi, q' \!=\! \tau(q,\sigma)\}.\label{next_state}
\end{align}
Additionally, a notion of \emph{labeling function} is required to connect a dt-CS to a given DFA $\mathcal{A}$.
\begin{definition}\label{def:sactisfaction_DFA} \emph{(Labeling Function)}
    Consider a dt-CS $\Sigma:=(X, X_0,$ $X_F, U,f, Y,h)$, a DFA $\mathcal{A}\!=\! (Q, q_0, \Pi,\tau,$ $ F)$, and a finite state run $\mathbf{x}_{x_0,\nu}:= (x(0),\ldots,x(H))\!\in\! X^{H+1}$ of $\Sigma$ with $H\!\in\!\mathbb{N}_{>0}$ under an input run $\nu: = (\nu(0),\ldots,\nu(H-1))$.
    We define a labeling function $L: X\rightarrow \Pi$  and a function $L_H:X^{H+1}\rightarrow \Pi^{H+1}$ such that the trace of $\mathbf{x}_{x_0,\nu}$ over $\Pi$ is $\sigma :=\!L_H(\mathbf{x}_{x_0,\nu})\!=\!$ $(\sigma_0,\sigma_1,\ldots,\sigma_{H})$ with $\sigma_k=L(\mathbf{x}_{x_0,\nu}(k))$ for all $k\in[0,H]$.  
\end{definition}

\subsection{Main Problem}
In this paper, our primary focus is on the verification of $\delta$-approximate $K$-diagnosability over a dt-CS. 
When the dt-CS is diagnosable, we also aim to construct a \emph{diagnoser} capable of detecting a fault within a maximum of $K$ time steps after its initial occurrence, without indicating a fault if none has occurred. 
This diagnoser, referred to as a \emph{($\delta$,$K$)-diagnoser}, is formally defined below.
\begin{definition}\label{def:diagnoser}
	\emph{(($\delta$,$K$)-diagnoser)} 
	Consider a dt-CS $\Sigma:=(X, X_0, X_F, U,f,Y,h)$, constants $\delta\in\mathbb{R}_{\geq 0}$ and $K\in\mathbb{N}$. 
	A \emph{($\delta$,$K$)-diagnoser} is a function
 \begin{equation}
     D: Y^{*} \rightarrow \{0,1\},\label{Dfunction}
 \end{equation} satisfying the following conditions:
	\begin{itemize}
		\item (C1) For all $k\in \mathbb{N}$ and $\mathbf{x}_{x_0,v}$ such that $\mathbf{x}_{x_0,v}(i)\notin X_F$, $\forall i\in[0,k-1]$, and $\mathbf{x}_{x_0,v}(k)\in X_F$, one has  $D(\mathbf{y}^{\delta}_{x_0,\nu}(0),\ldots,\mathbf{y}^{\delta}_{x_0,\nu}(k+K))=1$;
		\item (C2) For all $\mathbf{x}_{x_0,v}$ such that $\forall i\in \mathbb{N}$, $\mathbf{x}_{x_0,v}(i)\notin X_F$, one has $D(\mathbf{y}^{\delta}_{x_o,\nu})=0$,
	\end{itemize}
    where $\mathbf{y}^{\delta}_{x_0,\nu} \in Y^*$, with  $\mathbf{y}^{\delta}_{x_0,\nu}$ being an imprecise observation defined as in~\eqref{imprecise}. 
\end{definition}

Having the definition of a ($\delta$,$K$)-diagnoser, we are ready to formulate the main problem to be tackled in this paper.
\begin{problem}\label{main_prob}
		Consider a dt-CS $\Sigma:=(X, X_0,X_F, U,f,Y,h)$ as in Definition \ref{def:sys1}, constants $\delta\in\mathbb{R}_{\geq 0}$, and $K\in\mathbb{N}$. 
		\begin{itemize}
			\item Verify whether or not $\Sigma$ is $\delta$-approximate $K$-diagnosable as described in Definition~\ref{def:diagnosability};
			\item If $\Sigma$ is $\delta$-approximate $K$-diagnosable, then design a ($\delta$,$K$)-diagnoser as described in Definition~\ref{def:diagnoser}. 
		\end{itemize}
\end{problem}

To better illustrate the theoretical results of this paper, the following dt-CS with finite state and input sets will be leveraged throughout the paper.
Note that the results proposed in this paper can be applied to dt-CS with both continuous and discrete state and input sets.
A case study with continuous state and input sets will be used to showcase the effectiveness of the propose results (cf. Section~\ref{sec:case}).
\begin{example}
	(running example) As a running example, we focus on a dt-CS $\Sigma_{\text{run}}:=(X, X_0,X_F, U,f,Y,h)$ as demonstrated in Figure~\ref{runningexample}. 
    Concretely, one has 
    $X = \{0, 1.2, 2.2, 3.2, 4.2, 5.2, 6.2, 7.2, 9\}$, 
    $X_0 = \{ 0\}$,
    $X_F = \{1.2\}$,
    $U=\{1,2\}$,
    $Y = X$, 
    and $h:X \rightarrow Y$ is defined such that $y=x$ for any $x\in X$.
    For the running example, we are interested in two $\delta$-approximate $K$-diagnosability properties:
    1) $\delta = 1$, $K = 2$; 2) $\delta = 1$, $K = 3$.  
\end{example}
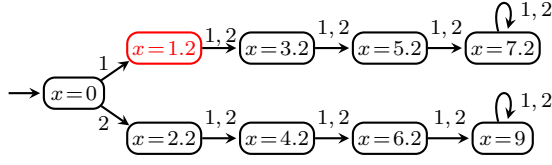
\begin{figure} 
	\centering
	\centering
	\begin{tikzpicture}[->,>=stealth,shorten >=1pt,auto,node distance=1.5cm,
		thick,base node/.style={rectangle, align = center, draw, minimum height=3mm,minimum width=8mm, rounded corners =1.5mm, font=\footnotesize},]
		
		
		\node[initial,initial text={}, base node] (1) at (0,1) {$\! x \! =\!0 \!$};
		\node[state, base node, red](2) at (1.2, 1.6) {$\! x \! =\! 1.2\!$};
		\node[state, base node](3) [right of= 2] {$\! x \! =\! 3.2\!$};
		\node[state, base node](4) [right of= 3] {$\! x \! =\! 5.2\!$};
		\node[state, base node](5) [right of= 4] {$\! x \! =\! 7.2\!$};
		
		\node[state, base node](6) at (1.2, 0.4) {$\! x \! =\! 2.2\!$};
		\node[state, base node](7) [right of= 6] {$\! x \! =\! 4.2\!$};
		\node[state, base node](8) [right of= 7] {$\! x \! =\! 6.2\!$};
		\node[state, base node](9) [right of= 8] {$\! x \! =\! 9\!$};

		
		\path ($(1.north east)+(-0.05,-0.05)$) edge node[left=0.8mm,pos=0.8] {\fontsize{8}{1}$1$} ($(2.south west)+(0.07,0.07)$)
		($(1.south east)+(-0.05,0.05)$) edge node[left=0.8mm,pos=0.8] {\fontsize{8}{1}$2$} ($(6.north west)+(0.07,-0.07)$);
		
		\path (2) edge node[yshift=-0.1cm] {\fontsize{8}{1}$1,2$} (3)
		(3) edge node {\fontsize{8}{1}$1,2$} (4)
		(4) edge node {\fontsize{8}{1}$1,2$} (5);
		
		\path (6) edge node {\fontsize{8}{1}$1,2$} (7)
		(7) edge node {\fontsize{8}{1}$1,2$} (8)
		(8) edge node {\fontsize{8}{1}$1,2$} (9);
		
		\path[] (5) edge [loop above] node [right,xshift=0.1cm,yshift=-0.1cm] {\fontsize{8}{1}$1,2$} (5)
		(9) edge [loop above] node [right,xshift=0.1cm,yshift=-0.1cm] {\fontsize{8}{1}$1,2$} (9);
	\end{tikzpicture}
		\caption{A finite state system as a running example, with the state in red being the faulty one.}\label{runningexample}
	\end{figure}

\section{Verification of $\delta$-Approximate $K$-Diagnosability Properties}\label{sec3}
\subsection{Verification via ($\delta$, $K$)-Deterministic Finite Automata}\label{sec3.1}

Consider a dt-CS $\Sigma\!=\!(X,X_0,X_F,U,f,Y,h)$. 
To tackle Problem~\ref{main_prob}, we introduce an \emph{augmented system} associated with $\Sigma$, which is the product between $\Sigma$ and itself, defined as
\begin{align}
    \Sigma_{\text{aug}}:= (X \times X, X_0  \times X_0, X_F \times X_F,U \times U,f \times f, Y  \times Y, h \times h).\label{aug_sys}
\end{align}
Here, we denote by $(x,\hat x) \!\in\! X \!\times\! X$ a state pair of  $\Sigma \!\times \!\Sigma$, and by $(\mathbf{x}_{x_0,\nu},  \hat{\mathbf{x}}_{\hat x_0,\hat \nu})$ the state run of $\Sigma \times \Sigma$ starting from $(x_0, \hat x_0)$ under input run ($\nu, \hat \nu$).
Moreover, we use $\mathcal{R}\!=\!X \!\times X$ to represent the augmented state set.
Then, considering the desired $\delta$-approximate $K$-diagnosability property, we propose the construction of a so-called ($\delta$,$K$)-DFA over the augmented system $\Sigma_{\text{aug}}$, which plays a key role in our main results for the verification over the original dt-CS  $\Sigma$ against the desired diagnosability property.
\begin{definition}\label{dKDFA}
	\emph{(($\delta$,$K$)-DFA)}
	Consider a dt-CS $\Sigma:=(X, X_0,X_F,U,f, Y,h)$, the corresponding augmented system $\Sigma_{\text{aug}}$ as in~\eqref{aug_sys}, and constants $\delta \in\mathbb{R}_{>0}$, $K \in\mathbb{N}$. Consider sets $\mathcal{P}_1:=\{(x,\hat{x}):||h(x)-h(\hat{x})||\leq \delta\}$, $\mathcal{P}_2:=\{(x,\hat{x}):x\in X_F, \hat{x}\in X\}$, and $\mathcal{P}_3:=\{(x,\hat{x}):\hat{x}\in X_F, x \in X\}$, with $\mathcal{P}_1,\mathcal{P}_2,\mathcal{P}_3\subseteq \mathcal{R}$.
	We define a ($\delta$,$K$)-DFA for the $\delta$-approximate $K$-diagnosability property, denoted by $\mathcal{A}_{(\delta,K)}\!=\! (\bar{Q},\bar{q}_0, \bar{\Pi},\bar{\tau},\bar{F})$, and a labeling function $\bar{L}:\mathcal{R}\rightarrow \bar{\Pi}$, in which 
	\begin{itemize}
		\item State set $\bar{Q}:=\{\bar{q}_0,q_1,\ldots,q_K,\bar{F},q_{\text{trap}}\}$.
		\item Alphabet $\bar{\Pi}:=\{\sigma_1,\sigma_2,\sigma_3\}$, with $\bar{L}$ being defined such that 
  $\bar{L}^{-1}(\sigma_1):= \mathcal{P}_1 \cap (\mathcal{R}\setminus \mathcal{P}_2) \cap (\mathcal{R}\setminus \mathcal{P}_3)$,
  $\bar{L}^{-1}(\sigma_2):= (\mathcal{R}\setminus \mathcal{P}_1) \cup (\mathcal{P}_1 \cap \mathcal{P}_3)$, 
  and $\bar{L}^{-1}(\sigma_3):= \mathcal{P}_1\cap \mathcal{P}_2\cap(\mathcal{R}\setminus \mathcal{P}_3)$.  
		\item Transition function $\bar{\tau}$ is defined such that 
		$q_{0} = \bar{\tau}(q_0,\sigma_1)$; 
		$q_{1} = \bar{\tau}(q_0,\sigma_3)$;
		$q_{\text{trap}} = \bar{\tau}(q_i,\sigma_2)$, $\forall i\in[0,K]$;
		$q_{i+1} = \bar{\tau}(q_i,\sigma_1)$, and $q_{i+1} = \bar{\tau}(q_i,\sigma_3)$, $\forall i\in[1,K-1]$;
		$\bar{F} = \bar{\tau}(q_K,\sigma_1)$ and $\bar{F} = \bar{\tau}(q_K,\sigma_3)$;
		$q_{\text{trap}} = \bar{\tau}(q_{\text{trap}},\sigma_j)$ and $\bar{F} = \bar{\tau}(\bar{F},\sigma_j)$, for all $j\in\{1,2,3 \}$.  
	\end{itemize}
\end{definition}

For simple presentation, $\mathcal{A}_{(\delta,K)}$ is referred to as \emph{($\delta$,$K$)-DFA} in the rest of the paper.
Now, let us consider the running example to construct the ($\delta$,$K$)-DFA with respect to the desired diagnosability properties.

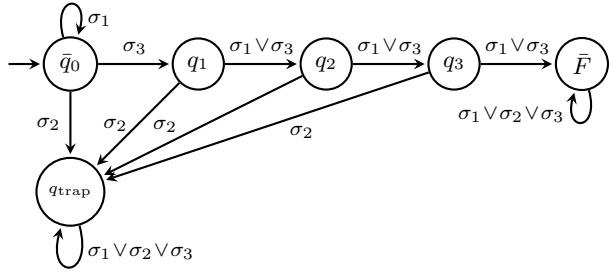
\begin{figure} 
	\centering
	\centering
	\begin{tikzpicture}[->,>=stealth,shorten >=1pt,auto,node distance=1.7cm,
		thick,base node/.style={circle,draw,minimum size=0.5mm, font=\small},]
		
		
		\node[initial,initial text={}, base node] (q0) at (0,1) {$\bar{q}_0$};
		\node[state, base node](q1) [right of= q0] {$q_1$};
		\node[state, base node](q2) [right of= q1] {$q_2$};
		\node[state, base node](q3) [right of= q2] {$q_3$};
		\node[state, base node](F) [right of= q3] {$\bar{F}$};
		
		\node[state, base node, font=\tiny](tr) [below of= q0] {$q_{\text{trap}}$};

		\path (q0) edge node {\fontsize{8}{1}$\sigma_3$} (q1)
		(q1) edge node {\fontsize{8}{1}$\sigma_1  \! \vee  \! \sigma_3$} (q2)
		(q2) edge node {\fontsize{8}{1}$\sigma_1  \! \vee  \! \sigma_3$} (q3)
		(q3) edge node {\fontsize{8}{1}$\sigma_1  \! \vee  \! \sigma_3$} (F);
		
		\path (q0) edge node[left] {\fontsize{8}{1}$\sigma_2$} (tr)
		(q1) edge node[left] {\fontsize{8}{1}$\sigma_2$} (tr)
		(q2) edge node[left=0.2] {\fontsize{8}{1}$\sigma_2$} (tr)
		(q3) edge node[right=0.4,pos=0.55] {\fontsize{8}{1}$\sigma_2$} (tr);
		
		\path[] (q0) edge [loop above] node [right,xshift=0.1cm, yshift=-0.2cm] {\fontsize{8}{1}$\sigma_1$} (q0);
		\path[] (tr) edge [loop below] node [right,xshift=0.1cm, yshift=0.2cm] {\fontsize{8}{1}$\sigma_1 \! \vee  \! \sigma_2  \! \vee  \! \sigma_3$} (tr);
		\path[] (F) edge [loop below] node [left=0.1,yshift=0.1cm] {\fontsize{8}{1}$\sigma_1  \! \vee  \! \sigma_2 \! \vee \! \sigma_3$} (F);
		
	\end{tikzpicture}
		\caption{The $(1,3)$-DFA for the running example.}\label{DFA}
	\end{figure}

\addtocounter{example}{-1}
\begin{example}[continued]
Consider a $\delta$-approximate $K$-diagnosability property with $\delta=1$ and $K=3$. The corresponding ($\delta$,$K$)-DFA is shown in Figure~{\ref{DFA}}, with the alphabet $\Pi := \{\sigma_1, \sigma_2, \sigma_3\}$ being defined such that 
    $\bar{L}^{-1}(\sigma_1):= \{(x,\hat{x}):||h(x)-h(\hat{x})||\leq 1, x\neq 1.2,\hat{x}\neq 1.2\}$,
    $\bar{L}^{-1}(\sigma_2):= \{(x,\hat{x}):||h(x)-h(\hat{x})||> 1\} \cup \{(x,\hat{x}):||h(x)-h(\hat{x})||\leq 1, \hat{x}= 1.2\}$,
    and $\bar{L}^{-1}(\sigma_3):= \{(x,\hat{x}):||h(x)-h(\hat{x})||\leq 1, x= 1.2, \hat{x}\neq 1.2\}$.
\end{example}

To verify $\delta$-approximate $K$-diagnosability of a dt-CS leveraging the corresponding ($\delta$,$K$)-DFA, a product system between the augmented system $\Sigma_{\text{aug}}$ as in~\eqref{aug_sys} and the ($\delta$,$K$)-DFA $\mathcal{A}_{(\delta,K)}$ is constructed, which is defined as follows.
\begin{definition}\label{def:product_dtCS} 
	Consider a dt-CS $\Sigma:=(X, X_0,X_F, U, f, Y,$ $h)$, the corresponding augmented system $\Sigma_{\text{aug}}$ as in~\eqref{aug_sys}, and a ($\delta$,$K$)-DFA $\mathcal{A}_{(\delta,K)}=(\bar{Q},\bar{q}_0, $ $\bar{\Pi},\bar{\tau},\bar{F})$ with a labeling function $\bar{L}: \mathcal{R}\rightarrow \Pi$ as in Definition~\ref{dKDFA}.
	The product system between $\Sigma_{\text{aug}}$ and $\mathcal{A}_{(\delta,K)}$ is a tuple
	\begin{equation}
		\Sigma_{\text{aug}} \otimes \mathcal{A}_{(\delta,K)} := (\tilde{X}, \tilde{X}_0,\tilde{U},\tilde{f}), \label{product}
	\end{equation}
	where
	$\tilde{X}:=\mathcal{R}\times \bar{Q}$ is the state set;
	$\tilde{X}_0$ is the initial state set, with $\tilde{x}_0:=(x_0,\hat{x}_0,\tilde{q}_0)\in\tilde{X}_0$, $(x_0,\hat{x}_0)\in X_0 \times X_0$, and
		$\tilde{q}_0 = \bar{\tau}(\bar{q}_0,\bar{L}(x_0,\hat{x}_0))$; 
	$\tilde{U}:= U\times U$ is the input set; 
	and $\tilde{f}:\tilde{X}\times \tilde{U}\rightarrow \tilde{X}$ is the transition function such that for any $\tilde{x}:=(x,\hat{x},q)\in\tilde{X}$ and $\tilde{u}:=(u,\hat{u})$, one has $\tilde{x}_{+}=\tilde{f}(\tilde{x},\tilde{u}):=(x_+,\hat{x}_+,q_+)$, with $x_+ = f(x,u)$, $\hat{x}_+ = f(\hat{x},\hat{u})$, and $q_+ = \bar{\tau}(q,\bar{L}(x_+,\hat{x}_+))$. 
\end{definition}

Similarly, $\tilde{\nu}: = (\tilde{\nu}(0),\ldots,\tilde{\nu}(k),\ldots)$ denotes an input run of $\Sigma_{\text{aug}} \otimes \mathcal{A}_{(\delta,K)}$, and
$\tilde{\mathbf{x}}_{\tilde{x}_0,\tilde{\nu}}:= (\tilde{x}(0),\ldots,\tilde{x}(k),\ldots)$ denotes a state run of $\Sigma_{\text{aug}} \otimes \mathcal{A}_{(\delta,K)}$ initialized from $\tilde{x}_0$ under input run $\tilde{\nu}$, i.e., $\tilde{x}(0)=\tilde{x}_0$, $\tilde{x}(k+1)=\tilde{f}(\tilde{x}(k),\tilde{\nu}(k))$, $\forall k\in \mathbb{N}$.

Having the product system $\Sigma_{\text{aug}} \otimes \mathcal{A}_{(\delta,K)}$ as in~\eqref{product}, we are ready to propose one of the main results of this paper regarding how to verify whether or not a dt-CS is $\delta$-approximate $K$-diagnosable.
\begin{resp}
	\begin{theorem}\label{thm:1}
		Consider a dt-CS $\Sigma\!=\!(X,X_0,X_F,U,f,Y,h)$ as in Definition \ref{def:sys1}, constants $\delta \in\mathbb{R}_{>0}$, $K \in\mathbb{N}$, the corresponding augmented system $\Sigma_{\text{aug}}$ as in~\eqref{aug_sys}, the ($\delta$,$K$)-DFA $\mathcal{A}_{(\delta,K)}=(\bar{Q},\bar{q}_0, \bar{\Pi},\bar{\tau},\bar{F})$ with a labeling function $\bar{L}: \mathcal{R}\rightarrow \Pi$ as in Definition~\ref{dKDFA}, and the product system $\Sigma_{\text{aug}} \otimes \mathcal{A}_{(\delta,K)}$ as in~\eqref{product}.
		The dt-CS $\Sigma$ is $\delta$-approximate $K$-diagnosable, if and only if $\nexists  \tilde{\mathbf{x}}_{\tilde{x}_0,\tilde{\nu}}$ 
  such that $\tilde{\mathbf{x}}_{\tilde{x}_0,\tilde{\nu}}(k)\in \mathcal{R}\times \bar{F}$ for some $k\in\mathbb{N}$.
	\end{theorem}
\end{resp}
The proof of Theorem~\ref{thm:1} is provided in the Appendix.
As a key insight, Theorem~\ref{thm:1} converts the desired diagnosability property over the dt-CS to a safety property over the product system $\Sigma_{\text{aug}}\otimes \mathcal{A}$ as in~\eqref{product}.
Accordingly, one can verify the $\delta$-approximate $K$-diagnosability property by verifying whether or not the \textit{``unsafe set''} $\mathcal{R}\times \bar{F}$ can be reached by the product system $\Sigma_{\text{aug}}\otimes \mathcal{A}$.
Next, we introduce a corollary to establish the conditions under which one can infer that the dt-CS of interest is \emph{not} diagnosable.
\begin{corollary}\label{cor1}
		Consider a dt-CS $\Sigma\!=\!(X,X_0,X_F,U,f,Y,h)$ as in Definition \ref{def:sys1} and constants $\delta \in\mathbb{R}_{>0}$, $K \in\mathbb{N}$.
		The dt-CS $\Sigma$ lacks $\delta$-approximate $K$-diagnosability if and only if $ \exists \tilde{\mathbf{x}}_{\tilde{x}_0,\tilde{\nu}} $ 
  such that $\tilde{\mathbf{x}}_{\tilde{x}_0,\tilde{\nu}}(k)\in \mathcal{R}\times \bar{F}$ for some $k\in\mathbb{N}$, with $\tilde{\nu}$ and $\tilde{\mathbf{x}}_{\tilde{x}_0,\tilde{\nu}}$ being some input  and state runs generated by the product system $\Sigma_{\text{aug}} \otimes \mathcal{A}_{(\delta,K)}$ in~\eqref{product}, respectively.  
	\end{corollary}
 
Note that Corollary~\ref{cor1} is a direct result of Theorem~\ref{thm:1}, and we omit its proof for simplicity.
So far, we have introduced conditions under which a system is (not) $\delta$-approximate $K$-diagnosable.
In Section~\ref{sec3.2}, we provide sufficient results to check these conditions via a notion of \emph{hybrid barrier certificates} over the product system $\Sigma_{\text{aug}} \otimes \mathcal{A}_{(\delta,K)}$. 
This notion was inspired by the notion of barrier in~\cite{murali2023co}.

\subsection{Verifying (the Lack of) Approximate $K$-Diagnosability}\label{sec3.2}
In the previous subsection, we reformulated the verification of (the lack of) diagnosability as a safety verification problem over the product system $\Sigma_{\text{aug}} \otimes \mathcal{A}_{(\delta,K)}$ in~\eqref{product}.
To solve this safety verification problem, in this subsection, we deploy a notion of \emph{hybrid barrier certificates} (HBC) over the \emph{hybrid} state set of $\Sigma_{\text{aug}} \otimes \mathcal{A}_{(\delta,K)}$.
The notion of hybrid barrier certificates was introduced for the first time in~\cite{murali2023co}, but in a different context.
First, we propose the following result related to the verification of $\delta$-approximate $K$-diagnosability.
\begin{resp}
	\begin{theorem}\label{veri_Diag}
		Consider a dt-CS $\Sigma\!=\!(X,X_0,X_F,U,f,Y,h)$ in Definition \ref{def:sys1}, constants $\delta \in\mathbb{R}_{>0}$, $K \in\mathbb{N}$, the ($\delta$,$K$)-DFA $\mathcal{A}_{(\delta,K)}=(\bar{Q},\bar{q}_0, \bar{\Pi},\bar{\tau},\bar{F})$ with a labeling function $\bar{L}: \mathcal{R}\rightarrow \Pi$ in Definition~\ref{dKDFA}, and the product system $\Sigma_{\text{aug}} \otimes \mathcal{A}_{(\delta,K)}= (\tilde{X}, \tilde{X}_0,\tilde{U},\tilde{f})$ in~\eqref{product}.
		If there exists a function $\mathcal{B}: \tilde{X}\rightarrow \mathbb{R}$ 
  such that
		\begin{align}
			&\forall \tilde{x} \in \tilde{X}_0,\quad\quad\quad\quad\ \quad\quad \quad\ \ \,\mathcal{B}(\tilde{x})\leq 0;\label{cd1_1}\\
			&\forall \tilde{x} \in \mathcal{R}\times \bar{F}, \quad\quad\quad\quad\quad \quad \ \ \mathcal{B}(\tilde{x})>0;\label{cd1_2}\\
			&\forall \tilde{x} \!\in\! \mathcal{R}\!\times\! \bar{Q}\setminus \{q_{\text{trap}}\}, \forall \tilde{u}\!\in\! \tilde{U},\ \mathcal{B}(\tilde{f}(\tilde{x},\tilde{u}))\!\leq\! \mathcal{B}(\tilde{x}),\label{cd1_3}
		\end{align}
	then $\nexists \tilde{\mathbf{x}}_{\tilde{x}_0,\tilde{\nu}} $ 
 such that $\tilde{\mathbf{x}}_{\tilde{x}_0,\tilde{\nu}}(k)\in \mathcal{R}\times \bar{F}$ for some $k\in\mathbb{N}$, with $\tilde{\mathbf{x}}_{\tilde{x}_0,\tilde{\nu}}$ being any state run that can be generated by $\Sigma_{\text{aug}} \otimes \mathcal{A}_{(\delta,K)}= (\tilde{X}, \tilde{X}_0,\tilde{U},\tilde{f})$ under an input run $\tilde{\nu}$. 
	\end{theorem}
\end{resp}
The proof of Theorem~\ref{veri_Diag} is provided in the Appendix.
As a key insight,~\eqref{cd1_3} guarantees that $\mathcal{B}(x,\hat{x},q)$ is not increasing along any state run of $\Sigma_{\text{aug}} \otimes \mathcal{A}_{(\delta,K)}$ within the set $\mathcal{R}\times \bar{Q}\setminus \{q_{\text{trap}}\}$. 
Therefore, considering~\eqref{cd1_1} and~\eqref{cd1_2}, any state run starting from the set $\Tilde{X}_0$ will never reach the set $\mathcal{R}\times \bar{Q}$. 
Hence, according to Theorem~\ref{thm:1}, the existence of an HBC indicates that the dt-CS $\Sigma$ is $\delta$-approximate $K$-diagnosable.
In the rest of the paper, a function $\mathcal{B}(x,\hat{x},q)$ satisfying conditions~\eqref{cd1_1}-\eqref{cd1_3} is referred to as \emph{$\mathcal{B}$-augmented-barrier-certificate} ($\mathcal{B}$-HBC) for the product system $\Sigma_{\text{aug}} \otimes \mathcal{A}_{(\delta,K)}$. 
Next, we propose another type of HBC, which can be used to verify the lack of $\delta$-approximate $K$-diagnosability.
\begin{resp}
	\begin{theorem}\label{veri_Lack_Diag}
		Consider a dt-CS $\Sigma\!=\!(X,X_0,X_F,U,f,Y,h)$ in Definition \ref{def:sys1}, constants $\delta \in\mathbb{R}_{>0}$, $K \in\mathbb{N}$, the ($\delta$,$K$)-DFA $\mathcal{A}_{(\delta,K)}=(\bar{Q},\bar{q}_0, \bar{\Pi},\bar{\tau},\bar{F})$ with a labeling function $\bar{L}: \mathcal{R}\rightarrow \Pi$ in Definition~\ref{dKDFA}, and the product system $\Sigma_{\text{aug}} \otimes \mathcal{A}_{(\delta,K)}= (\tilde{X}, \tilde{X}_0,\tilde{U},\tilde{f})$ in~\eqref{product}.
		Suppose that there exists a function $\mathcal{V}: \tilde{X}\rightarrow \mathbb{R}$ such that
		\begin{align}
			&\forall \tilde{x} \in \tilde{X}_0,\quad\quad\quad\quad\quad\quad\quad\ \ \,\mathcal{V}(\tilde{x})\leq 0;\label{cd2_1}\\
			&\forall \tilde{x} \!\in\! (\tilde{P}(\mathcal{R})\setminus \mathcal{R})\! \times\! (\bar{Q}\setminus\bar{F}),\ \mathcal{V}(\tilde{x}) > 0;\label{cd2_2}\\
			&\forall \tilde{x} \in \overline{\mathcal{R}}\times (\bar{Q}\setminus \bar{F}), \exists\tilde{u}\in \tilde{U}, \ \mathcal{V}(\tilde{f}(\tilde{x},\tilde{u}))\!< \!\mathcal{V}(\tilde{x}),\label{cd2_3}
		\end{align}
		where $\tilde{P}(\mathcal{R}):=\{(x,\hat{x})~|~ \exists u,\hat{u}\in U, x' = f(x,u), \hat{x}'=f(\hat{x},\hat{u}), \text{ where }(x',\hat{x}')\in\mathcal{R}\}$. Then 
        there exists an input run $\tilde{\nu}$ and a state run $\tilde{\mathbf{x}}_{\tilde{x}_0,\tilde{\nu}}$ such that $\tilde{\mathbf{x}}_{\tilde{x}_0,\tilde{\nu}}(k)\in \mathcal{R}\times \bar{F}$ for some $k\in\mathbb{N}$.
	\end{theorem}
\end{resp}
The proof of Theorem~\ref{veri_Lack_Diag} is provided in the Appendix.
As a key insight,~\eqref{cd2_3} ensures that $\mathcal{V}(x,\hat{x},q)$ decreases strictly along the state run of $\Sigma_{\text{aug}}\otimes \mathcal{A}_{\delta,K}$ within the set $\overline{\mathcal{R}}\times (\bar{Q}\setminus \bar{F})$. 
Therefore, any state run starting from the set $\overline{\mathcal{R}}\times (\bar{Q}\setminus \bar{F})$ will eventually leave $\overline{\mathcal{R}}\times (\bar{Q}\setminus \bar{F})$ since $\overline{\mathcal{R}}\times (\bar{Q}\setminus \bar{F})$ is bounded. 
On the other hand,~\eqref{cd2_1} and \eqref{cd2_2} ensure that any state run starting from the set $\tilde{X}_0$ does not enter the set $(\mathbb{R}^n\backslash \overline{\mathcal{R}})\times (\bar{Q}\setminus \bar{F})$. Therefore, the existence of a function $\mathcal{V}(x,\hat{x},q)$ satisfying \eqref{cd2_1}-\eqref{cd2_3} indicates that any state run initialized from the set $\tilde{X}_0$ would eventually leave the set $\overline{\mathcal{R}}\times (\bar{Q}\setminus \bar{F})$ by entering the set $\mathcal{R}\times \bar{F}$.
Therefore, by applying Corollary~\ref{cor1}, one can readily verify that the existence of a function $\mathcal{V}$ implies that the dt-CS $\Sigma$ is not $\delta$-approximate $K$-diagnosable.
Hereafter, a function $\mathcal{V}(x,\hat{x},q)$ satisfying~\eqref{cd2_1}-\eqref{cd2_3} is referred to as \emph{$\mathcal{V}$-augmented-barrier-certificate} ($\mathcal{V}$-HBC) for the product system $\Sigma_{\text{aug}} \otimes \mathcal{A}_{(\delta,K)}$.

To sum up, by applying Theorem~\ref{veri_Diag} (resp. Theorem~\ref{veri_Lack_Diag}), the verification of (resp. the lack of) diagnosability can be solved by computing a $\mathcal{B}$-HBC (resp. $\mathcal{V}$-HBC) for the product system $\Sigma_{\text{aug}} \otimes \mathcal{A}_{(\delta,K)}$. 
In the next section, we discuss the computation of both HBCs in detail.

\section{Computation of Hybrid Barrier Certificates}\label{sec4}
In this section, we propose two methodologies to compute $\mathcal{B}$-HBC and $\mathcal{V}$-HBC as introduced in Theorems~\ref{veri_Diag} and \ref{veri_Lack_Diag}, respectively. 
The first methodology deploys a CEGIS framework~\cite{Lezama_2008_thesis}, under which the HBCs are computed using Satisfiability Modulo Theories (SMT) and \texttt{Gurobi}~\cite{gurobi} solvers.
As the second approach, we cast the computation of the hybrid barrier certificates as a sum-of-squares (SOS) programming that can be solved using semi-definite-programming (SDP) solvers (e.g., Mosek~\cite{ApS2019MOSEK}).
Note that the notions of \emph{monomial}, \emph{polynomial}, and \emph{SOS polynomials} will be deployed when introducing the SOS-based computation approaches.
Here, we refer readers to~\cite{JarvisWloszek2005Control} for the definitions of these notions, and we omit them here due to the lack of space.

\subsection{Computing $\mathcal{B}$-HBC for Verifying Diagnosability}
Here, we focus on computing $\mathcal{B}$-HBC as in Theorem~\ref{veri_Diag}.
Concretely, given a ($\delta$,$K$)-DFA $\mathcal{A}_{(\delta,K)}\!=\! (\bar{Q},\bar{q}_0, \bar{\Pi},\bar{\tau},\bar{F})$ with a labeling function $\bar{L}: \mathcal{R}\rightarrow \Pi$ as in Definition~\ref{dKDFA} related to the desired diagnosability property, we consider the template of $\mathcal{B}$-HBC as
\begin{align}\label{eq:2}
	\mathcal{B}(x,\hat{x}, q):=\begin{cases}
		\mathcal{B}_0(x,\hat{x})\!:=\! \!\sum_{\mathsf{a}=1}^{z_0}c_{\mathsf{a},0}p_{\mathsf{a},0}(x,\hat{x}), \,\mbox{if}\, q \!=\! \bar{q}_0;\\
        \mathcal{B}_{\mathsf{b}}(x,\hat{x}) \!:= \!\!\sum_{\mathsf{a}=1}^{z_{\mathsf{b}}}\!c_{\mathsf{a},\mathsf{b}}p_{\mathsf{a},\mathsf{b}}(x,\hat{x}), \,\mbox{if}\, q \!=\! q_{\mathsf{b}},\mathsf{b}\!\in\![1,K];\\
        \mathcal{B}_{K+1}(x,\hat{x}) \!:= \! \!\sum_{\mathsf{a}=1}^{z_{K\!+\!1}}\!c_{\mathsf{a},K\!+\!1}\,p_{\mathsf{a},K\!+\!1}(x,\hat{x}), \,\mbox{if}\, q\! = \!\bar{F};\\
        \mathcal{B}_{K+2}(x,\hat{x}) \!:=\!\! \sum_{\mathsf{a}=1}^{z_{K\!+\!2}}\!c_{\mathsf{a},K\!+\!2}\,p_{\mathsf{a},K\!+\!2}(x,\hat{x}), \,\mbox{if}\, q \!=\! q_{\text{trap}};
	\end{cases}\!
\end{align} 
with $z_0,z_{\mathsf{b}},z_{K+1},z_{K+2}\in\mathbb{N}_{>0}$ being some known constants,
such that for all $\mathsf{r}\in[0,K+2]$, $\mathcal{B}_{\mathsf{r}}(x,\hat{x})$ is a polynomial function of a fixed degree with monomials $p_{1,\mathsf{r}}(x,\hat{x}),\ldots,p_{z_{\mathsf{r}},\mathsf{r}}(x,\hat{x})$ that are given, and coefficients $c_{1,\mathsf{r}},\ldots,c_{z_{\mathsf{r}},\mathsf{r}}$ which are unknown.
To improve the clarity of the presentation, we define a function 
\begin{align}
    \Delta:\bar{Q} \rightarrow [0,K+2],\label{delta}
\end{align}
that maps each state in $\bar{Q}$ to a natural number within $[0,K+2]$ according to~\eqref{eq:2}.
Furthermore, we define sets
\begin{align}
    &Q_{\text{init}} := \{ q\in \bar{Q}: \exists \sigma \in\bar{\Pi}, q=\bar{\tau}(\bar{q}_0,\sigma) \},\label{set1}\\
    & \mathcal{G}(q,q')\!:=\! \{(x,\hat{x})\!\in\! X\times X: \exists \sigma \!\in\! \bar{\Pi}, \bar{L}(x,\hat{x}) \!=\! \sigma, q' \!=\! \bar{\tau}(q,\sigma) \},\label{setg}\\
    &\bar{X}(q,q'):=\{(x,\hat{x},u,\hat{u})\in X\times X\times U \times U: \big(f(x,u),f(\hat{x},\hat{u})\big)\in \mathcal{G}(q,q'), q,q'\in\bar{Q} \}.\label{set2}
\end{align} 
Intuitively, $Q_{\text{init}}$ denotes the subset of $\bar{Q}$ that can be reached from $\bar{q}_0$ within one transition, while $\mathcal{G}(q,q')$ and  $\bar{X}(q,q')$ denote the subset of $X\times X$ and $X\times X\times U \times U$ that can trigger a transition from $q$ to $q'$ considering the labeling function $\bar{L}$, respectively.
Then, given a dt-CS $\Sigma\!=\!(X,X_0,X_F,U,f,Y,h)$ and the template of $\mathcal{B}(x,\hat{x}, q)$ as in~\eqref{eq:2}, one can rewrite~\eqref{cd1_1}-\eqref{cd1_3} as~\eqref{rw11}-\eqref{rw13}:
\begin{align}
    & \forall q\in Q_{\text{init}}, \forall (x,\hat{x})\in \mathcal{G}(\bar{q}_0,q), \mathcal{B}_{\Delta(q)}(x,\hat{x})\leq 0 ;\label{rw11}\\
    & \forall (x,\hat{x})\in X\times X, \mathcal{B}_{\Delta(\bar{F})}(x,\hat{x})> 0;\label{rw12}\\
    & \forall q\in \bar{Q}\setminus \{q_{\text{trap}}\},\forall q'\in \text{Nxt}(q),  \forall (x,\hat{x}, u,\hat{u}) \in \bar{X}(q,q'),  \mathcal{B}_{\Delta(q')}(f(x,u),f(\hat{x},\hat{u}))\leq \mathcal{B}_{\Delta(q)}(x,\hat{x});\label{rw13}
\end{align}
with $\text{Nxt}(q)$ as in~\eqref{next_state}, 
sets $Q_{\text{init}}$, $\mathcal{G}(\bar{q}_0,q)$, and $\bar{X}(q,q')$ as in~\eqref{set1}-\eqref{set2}, respectively.
Before discussing how to compute a $\mathcal{B}$-HBC over dt-CS, we provide an example of a $\mathcal{B}$-HBC with the template as in~\eqref{eq:2} for the running example.

\addtocounter{example}{-1}
\begin{example}[continued]
    Consider a $\delta$-approximate $K$-diagnosability with $\delta=1$ and $K =3$. One can verify that the following function $\mathcal{B}$ satisfies conditions~\eqref{rw11}-\eqref{rw13}:
    \begin{small}
            \begin{align}
	&\mathcal{B}(x,\hat{x}, q):=\nonumber\\
	&\left\{\begin{aligned}
	&\!\!\!-\!0.2075\!+\! 0.2024x \!+\! 0.2024\hat{x} \!-\! 0.1001x^2\!- \! 0.1435x \hat{x} \!-\! 0.1001\hat{x}^2 \ +\!0.0248x^3\! -\! 0.0134 x^2  \hat{x}\!-\! 0.0134x  \hat{x}^2 \!+\! 0.0248\hat{x}^3, \!\text{ if } q \!=\! {q}_0;\\
        &1.3026 \!+\! 0.9963x \!+\! 0.9963\hat{x} \!-\! 0.5042x^2 \!+\! 0.5593x \hat{x} \!-\! 0.5042\hat{x}^2\!+\!  0.0362x^3 \!+\! 0.0305 x^2  \hat{x} + 0.0305x  \hat{x}^2 \!+\! 0.0362\hat{x}^3, \!\text{ if } q \!=\! {q}_1;\\
         &0.8301 \!+\! 0.7739x \!+\! 0.7739\hat{x} \!-\! 0.3744x^2 \!+\! 0.3956x \hat{x} \!-\! 0.3744\hat{x}^2  + 0.0262x^3\!+\! 0.0229 x^2  \hat{x} \!+\! 0.0229x  \hat{x}^2 \!+\! 0.0262\hat{x}^3, \!\text{ if } q \!=\! {q}_2;\\
        &0.5178 \!+\! 0.5382x \!+\! 0.5382\hat{x} \!-\! 0.2785x^2 \!+\! 0.2973x \hat{x} \!-\! 0.2785\hat{x}^2  + 0.0246x^3 \!+\! 0.0096 x^2  \hat{x} \!+\! 0.0096x  \hat{x}^2\! +\! 0.0246\hat{x}^3,\! \text{ if } q \!=\! {q}_3;\\
        &\!\!\!-\!0.6935 \!+\! 0.1888x \!+\! 0.1888\hat{x} \!-\! 0.1292x^2 \!+\! 0.0198x \hat{x} \!-\! 0.1292\hat{x}^2   -\!0.0254x^3 \!+\! 0.0224 x^2  \hat{x} \!+\! 0.0224x  \hat{x}^2  \!-\! 0.0254\hat{x}^3,\! \text{ if } q \!=\! {q}_{\text{trap}};\\
        &0.2008 \!+\! 0.1378x \!+\! 0.1378\hat{x} \!+\! 0.0424x^2 \!-\! 0.0884x \hat{x} \!+\! 0.0424\hat{x}^2   +\! 0.0374x^3 \!-\! 0.0295 x^2  \hat{x} \!-\! 0.0295x  \hat{x}^2 \!+\! 0.0374\hat{x}^3, \!\text{ if } q \!=\! \bar{F}.\\
	\end{aligned}\right.\label{eq:B1}
\end{align}
    \end{small}
which can be obtained by solving a linear programming problem over constraints~\eqref{rw11}-\eqref{rw13} as the system in Figure~\ref{runningexample} has finite state and input sets.
\end{example}

\subsubsection{CEGIS-based Approach}\label{c-based_method1}
To compute a $\mathcal{B}$-HBC using SMT solvers under the CEGIS framework, we first rewrite the conditions~\eqref{rw11}-\eqref{rw13} as~\eqref{finrw11}-\eqref{finrw13}, respectively:
\begin{align}
    & \Phi_1 = \bigwedge_{q\in Q_{\text{init}},(x,\hat{x})\in \mathcal{G}(\bar{q}_0,q)}\!\!\!\!\! \!\!\!\!\!\!\!\!\!\!\mathcal{B}_{\Delta(q)}(x,\hat{x})\leq 0;\label{finrw11}\\
    &\Phi_2 = \bigwedge_{(x,\hat{x})\in X\times X} \mathcal{B}_{\Delta(\bar{F})}(x,\hat{x}) > 0;\label{finrw12}\\
    &\Phi_3 = \!\!\!\!\!\!\!\!\!\!\!\!\!\!\!\bigwedge_{\substack{q\in \bar{Q}\setminus \{q_{\text{trap}}\}, q'\in \text{Nxt}(q),\\(x,\hat{x},u,\hat{u})\in \bar{X}(q,q') }}\!\!\!\!\!\!\!\!\!\!\!\!\!\!\!\!\!\!\!{B}_{\Delta(q')}(f(x,u),f(\hat{x},\hat{u})))\!\leq\! \mathcal{B}_{\Delta(q)}(x,\hat{x}),\label{finrw13}
\end{align}
with sets $Q_{\text{init}}$, $\mathcal{G}(\bar{q}_0,q)$, $\bar{X}(q,q')$, and $\text{Nxt}(q)$ in \eqref{set1}, \eqref{setg}, \eqref{set2} and \eqref{next_state}, respectively.
Then, we collect $N$ and $P$ points from $X\times X$ and $U\times U$, respectively, to obtain sets $\mathcal{D}_x:=\{(x_1,\hat{x}_1),\ldots,(x_N,\hat{x}_N)\}$ and $\mathcal{D}_u:=\{(u_1,\hat{u}_1),\ldots,(u_P,\hat{u}_P)\}$.
Having sets $\mathcal{D}_x$ and $\mathcal{D}_u$, we first compute a function $\mathcal{B}(x,\hat{x}, q)$ as in~\eqref{eq:2} such that for all $(x_j,\hat{x}_j)\in\mathcal{D}_x$ and $(u_i,\hat{u}_i)\in\mathcal{D}_u$, $\Phi_1 \wedge \Phi_2 \wedge \Phi_3$ is satisfied.
Note that in this case, the computation of such a function $\mathcal{B}(x,\hat{x}, q)$ boils down to a linear programming problem.
As a key insight,  when substituting the elements of the sets $\mathcal{D}_x$ and $\mathcal{D}_u$ into the constraints in~\eqref{finrw11} -\eqref{finrw13}, the resulting constraints are all linear with respect to the unknown coefficients of $\mathcal{B}$ in~\eqref{eq:2}.

Suppose one obtains a function $\mathcal{B}(x,\hat{x}, q)$ such that $\Phi_1 \wedge \Phi_2 \wedge \Phi_3$ is satisfied for all elements of the sets $\mathcal{D}_x$ and $\mathcal{D}_u$ (otherwise a function $\mathcal{B}(x,\hat{x}, q)$ with the given template satisfying~\eqref{rw11}-\eqref{rw13} does not exist).
To check whether the obtained function $\mathcal{B}(x,\hat{x}, q)$ is indeed a $\mathcal{B}$-HBC satisfying~\eqref{finrw11}-\eqref{finrw13}, one can search for counterexamples $(x_c,\hat{x}_c)\in X \times X$ and $(u_c,\hat{u}_c)\in U \times U$ such that $\neg (\Phi_1 \wedge \Phi_2 \wedge \Phi_3)$ is true.
To this end, one can deploy \texttt{Gurobi}~\cite{gurobi} solver, or encode $\neg (\Phi_1 \wedge \Phi_2 \wedge \Phi_3)$ as an SMT query and use solver z3~\cite{moura_2008_z3}, to search for $(x_c,\hat{x}_c)$ and $(u_c,\hat{u}_c)$.
In those cases where a counterexamples $(x_c,\hat{x}_c)$ and/or $(u_c,\hat{u}_c)$ are found, we add them to $\mathcal{D}_x$ and $\mathcal{D}_u$ and repeat the process.
Otherwise, the obtained $\mathcal{B}(x,\hat{x}, q)$ is a valid $\mathcal{B}$-HBC.
The CEGIS-based computation of $\mathcal{B}$-HBC as in Theorem~\ref{veri_Diag} is summarized in Algorithm~\ref{alg:CEGIS}.

\IncMargin{0.5em}
\begin{algorithm2e}[ht!]
	\DontPrintSemicolon
	\Indm 
	\KwIn{A dt-CS $\Sigma:=(X, X_0,X_F,U,f,Y,h)$, a ($\delta$,$K$)-DFA $\mathcal{A}_{(\delta,K)}\!=\! (\bar{Q},\bar{q}_0, \bar{\Pi},\bar{\tau},\bar{F})$ with a labeling function $\bar{L}: \mathcal{R}\rightarrow \Pi$ as in Definition~\ref{dKDFA}, a template of function $\mathcal{B}(x,\hat{x},q)$ as in~\eqref{eq:2}, and $\mathsf{i}_{\text{max}}$ as the maximal number of iteration.}
	\KwOut{A $\mathcal{B}$-HBC $\mathcal{B}(x,\hat{x},q)$ if the computation terminates successfully.}
	\Indp
    $\mathsf{i}=0$, build the sets $\mathcal{D}_x:=\{(x_1,\hat{x}_1),\ldots,(x_N,\hat{x}_N)\}$ and $\mathcal{D}_u:=\{(u_1,\hat{u}_1),\ldots,(u_P,\hat{u}_P)\}$.\\
    \While{1}
	{
        Compute the coefficients of $\mathcal{B}(x,\hat{x},q)$ such that~\eqref{finrw11}-\eqref{finrw13} holds for all elements in the sets $\mathcal{D}_x$ and $\mathcal{D}_u$.\\
        \eIf{$\mathcal{B}(x,\hat{x},q)$ cannot be found}{
            The barrier with the given template doesn't exist.\\
            Terminate unsuccessfully.
        }
        {
            Compute counterexamples $(x_c,\hat{x}_c)$ and $(u_c,\hat{u}_c)$ satisfying $\neg (\Phi_1 \wedge \Phi_2 \wedge \Phi_3)$ for the obtained $\mathcal{B}(x,\hat{x},q)$.\\
            \eIf{$(x_c,\hat{x}_c)$ and $(u_c,\hat{u}_c)$ cannot be found}{
            $\mathcal{B}(x,\hat{x},q)$ is a $\mathcal{B}$-HBC.\\
            Terminate successfully.
            }
            {
            $\mathcal{D}_x:=\mathcal{D}_x \cup \{ (x_c,\hat{x}_c)\}$, $\mathcal{D}_u:=\mathcal{D}_u \cup \{ (u_c,\hat{u}_c)\}$.\\
            $\mathsf{i}=\mathsf{i}+1$.
            }
        }
        \If{$\mathsf{i} > \mathsf{i}_{\text{max}}$}{
        Terminate unsuccessfully.
        }
    }
    \caption{CEGIS-based computation of $\mathcal{B}$-HBC as in Theorem~\ref{veri_Diag}.} 
	\label{alg:CEGIS}
\end{algorithm2e}
\DecMargin{0.5em}

\subsubsection{SOS-based Approach}\label{s-based-method1}
With additional assumptions over dt-CS, as stated below, one can also compute a $\mathcal{B}$-HBC using SOS-based approaches.
\begin{assumption}\label{ass1}
	Consider a dt-CS $\Sigma:=(X, X_0,X_F, U,f, Y,h)$  in Definition~\ref{def:sys1}.
	We assume that the transition function $f$ is a polynomial function in variables $x$ and $u$, and the output function $h$ is a polynomial function in variable $x$.
\end{assumption}
Additionally, the following definition is also required for proposing the SOS-based computation of a $\mathcal{B}$-HBC. 
\begin{assumption}\label{def:function}
    Consider a dt-CS $\Sigma:=(X, X_0,X_F, $ $U,f, Y,h)$ in Definition~\ref{def:sys1}, constants $\delta \in\mathbb{R}_{>0}$, $K \in\mathbb{N}$, and the ($\delta$,$K$)-DFA $\mathcal{A}_{(\delta,K)}=(\bar{Q},\bar{q}_0, \bar{\Pi},\bar{\tau},\bar{F})$ with a labeling function $\bar{L}: \mathcal{R}\rightarrow \Pi$ in Definition~\ref{dKDFA}.
    We assume that we are given
    vectors of polynomial functions $g_0(q)(x,\hat{x})$, $g_1(x,\hat{x})$, $g_u(u,\hat{u})$, and $g_2(q,q')(x,\hat{x})$ of appropriate dimensions for some $q,q'\in \bar{Q}$, where:
    \begin{align}
        &\mathcal{G}(\bar{q}_0,q)\subseteq \{(x,\hat{x})\in\ \mathbb{R}^{2n} ~|~ g_0(q)(x,\hat{x})\geq 0 \},\label{eq_bg}\\
        &\mathcal{R} \subseteq \{ (x,\hat{x})\in\ \mathbb{R}^{2n}  ~|~ g_1(x,\hat{x})\geq 0 \},\label{eq_bg1}\\
        & \bar{X}(q,q') \!\subseteq\! \{(x,\hat{x},u,\hat{u})\in  \mathbb{R}^{2(n+m)} ~|~ g_2(q,q')(x,\hat{x},u,\hat{u})\!\geq\! 0 \},\\
        &\tilde{U}\subseteq \{ (u,\hat{u})\in\ \mathbb{R}^{2m} ~|~ g_u(u,\hat{u})\geq 0 \},\label{eq_bgu}
    \end{align}
    in which all inequalities are component-wise. 
\end{assumption}

For simple presentation, we omit $(x,\hat{x})$ and $(x,\hat{x},u,\hat{u})$ as arguments of these functions in the rest of the paper when it is clear from the context.
Now, we propose the SOS-based computation of $\mathcal{B}$-HBC as in Theorem~\ref{veri_Diag}.
	\begin{lemma}\label{lemma1} 
		Consider a dt-CS $\Sigma:=(X, X_0,X_F, U,f, Y,h)$ such that Assumption~\ref{ass1} holds. Consider constants $\delta \in\mathbb{R}_{>0}$, $K \in\mathbb{N}$, the ($\delta$,$K$)-DFA $\mathcal{A}_{(\delta,K)}=(\bar{Q},\bar{q}_0, \bar{\Pi},\bar{\tau},\bar{F})$,
		and functions $g_0(q)$, $g_1$, $g_2(q,q')$, and $g_u$ in Assumption~\ref{def:function}, with $q,q'\in Q$.
        Suppose there exist a constant $\epsilon>0$ and a function $\mathcal{B}(x,\hat{x},q)$ in the form of~\eqref{eq:2} such that the following expressions are SOS polynomials:
		\begin{align}
			&\forall q\in Q_{\text{init}}, \quad -\mathcal{B}_{\Delta(q)}(x,\hat{x})-\lambda_0^\top(q)g_0(q);\label{soscond11}\\
			&\mathcal{B}_{\Delta(\bar{F})}(x,\hat{x})-\lambda_1^\top(x,\hat{x})g_1-\epsilon;\label{soscond12}\\
            &\forall q\in \bar{Q}\setminus \{q_{\text{trap}}\},q'\in \bar{Q}, -\mathcal{B}_{\Delta(q')}({f}(x,u),f(\hat{x},\hat{u})) +\mathcal{B}_{\Delta(q)}(x,\hat{x})-\lambda_u^\top\!(q)g_u-\lambda_2^\top\!(q) g_2(q,q'),\label{soscond13}
		\end{align}
        where for any $q\in\bar{Q}$, $\lambda_1$ and $\lambda_0(q)$ (resp. $\lambda_u(q)$ and $\lambda_2(q)$) are vectors of SOS polynomials in variables $x$ and $\hat{x}$ (resp. $x$, $\hat{x}$, $u$, and $\hat{u}$)  of appropriate sizes.
		Then, function $\mathcal{B}(x,\hat{x},q)$ satisfies~\eqref{cd1_1}-\eqref{cd1_3}. 
	\end{lemma}

The proof of Lemma~\ref{lemma1} is provided in the Appendix.
So far, we have proposed two approaches for computing $\mathcal{B}$-HBC in Theorem~\ref{veri_Diag}.
In the next subsection, we discuss how to compute $\mathcal{V}$-HBC in Theorem~\ref{veri_Lack_Diag} for the product system $\Sigma_{\text{aug}} \otimes \mathcal{A}_{(\delta,K)}$.

\vspace{-0.3cm}

\subsection{Computing $\mathcal{V}$-HBC  for Verifying the Lack of Diagnosability}
To find a $\mathcal{V}$-HBC in Theorem~\ref{veri_Lack_Diag}, given a ($\delta$,$K$)-DFA $\mathcal{A}_{(\delta,K)}\!=\! (\bar{Q},\bar{q}_0, \bar{\Pi},\bar{\tau},\bar{F})$ with a labeling function $\bar{L}: \mathcal{R}\rightarrow \Pi$ as in Definition~\ref{dKDFA} related to the desired diagnosability property, we consider $\mathcal{V}(x,\hat{x}, q)$ in the form of
\begin{align}\label{eq:V}
	&\mathcal{V}(x,\hat{x}, q):=\begin{cases}
		\mathcal{V}_0(x,\hat{x}):= \sum_{\mathsf{a}=1}^{\alpha_0}s_{\mathsf{a},0}b_{\mathsf{a},0}(x,\hat{x}), \mbox{ if } q = \bar{q}_0;\\
        \mathcal{V}_{\mathsf{b}}(x,\hat{x}):= \sum_{\mathsf{a}=1}^{\alpha_{\mathsf{b}}}s_{\mathsf{a},\mathsf{b}}b_{\mathsf{a},\mathsf{b}}(x,\hat{x}), \mbox{ if } q = q_{\mathsf{b}},\mathsf{b}\in[1,K];\\
        \mathcal{V}_{K+1}(x,\hat{x}):= \sum_{\mathsf{a}=1}^{\alpha_{K+1}}s_{\mathsf{a},K+1}b_{\mathsf{a},K+1}(x,\hat{x}), \mbox{ if } q = \bar{F};\\
        \mathcal{V}_{K+2}(x,\hat{x}):= \sum_{\mathsf{a}=1}^{\alpha_{K+2}}s_{\mathsf{a},K+2}b_{\mathsf{a},K+2}(x,\hat{x}), \mbox{ if } q = q_{\text{trap}};
	\end{cases}
\end{align}
where $\alpha_0,\alpha_{\mathsf{b}},\alpha_{K+1},\alpha_{K+2}\in\mathbb{N}_{>0}$ are some known constants,
such that for all $\mathsf{r}\in[0,K+2]$, $\mathcal{V}_{\mathsf{r}}(x,\hat{x})$ is a polynomial function of a fixed degree with monomials $b_{1,\mathsf{r}}(x,\hat{x}),\ldots,b_{\alpha_{\mathsf{r}},\mathsf{r}}(x,\hat{x})$ which are given, and coefficients $s_{1,\mathsf{r}},\ldots,s_{\alpha_{\mathsf{r}},\mathsf{r}}$ which are unknown.
Similarly, one can use the function $\Delta$ as in~\eqref{delta} to match each state in $\bar{Q}$ to a natural number within $[0,K+2]$ according to~\eqref{eq:V}.
Then, given a dt-CS $\Sigma\!=\!(X,X_0,X_F,U,f,Y,h)$ in Definition \ref{def:sys1} and the template of $\mathcal{V}(x,\hat{x}, q)$ as in~\eqref{eq:V}, we rewrite~\eqref{cd2_1}-\eqref{cd2_3} as in~\eqref{rw21}-\eqref{rw23}, respectively:
\begin{align}
    & \forall q\in Q_{\text{init}}, \forall (x,\hat{x})\in \mathcal{G}(\bar{q}_0,q), \mathcal{V}_{\Delta(q)}(x,\hat{x})\leq 0 ;\label{rw21}\\
    & \forall q\in \bar{Q}\setminus \bar{F}, \forall (x,\hat{x})\in\tilde{P}(\mathcal{R})\setminus \mathcal{R}, \mathcal{V}_{\Delta(q)}(x,\hat{x})> 0;\label{rw22}\\
    & \forall q \!\in\! \bar{Q}\!\setminus\! \bar{F} ,\forall (x,\hat{x})\!\in\! \overline{\mathcal{R}}, \exists q'\!\in\! \text{Nxt}(q),  \exists (x,\hat{x}, u,\hat{u}) \!\in\! \bar{X}(q,q'),  \mathcal{V}_{\Delta(q')}(f(x,u),f(\hat{x},\hat{u}))< \mathcal{V}_{\Delta(q)}(x,\hat{x}),\label{rw23}
\end{align}
where $q' = \bar{\tau}(q,\bar{L}((f(x,u),f(\hat{x},\hat{u}))))$, $Q_{\text{init}}$, $\mathcal{G}(\bar{q}_0,q)$, and $\text{Nxt}(q)$ are as in~\eqref{set1}, \eqref{setg}, and \eqref{next_state}, respectively.

Before proceeding with discussing the systematic computation of a $\mathcal{V}$-HBC over dt-CS with continuous state and input sets, we revisit the running example to provide an example of a $\mathcal{V}$-HBC with the template as in~\eqref{eq:V}.
\addtocounter{example}{-1}
\begin{example}[continued]
    Consider a $\delta$-approximate $K$-diagnosability with $\delta=1$ and $K =2$.
    The following function $\mathcal{V}$ satisfies conditions~\eqref{rw21}-\eqref{rw23}:
            \begin{align*}
	&\mathcal{V}(x,\hat{x}, q):=\nonumber\\
	&\left\{\begin{aligned}
	 &\!-\!9.88\times 10^{-5}\!-\! 1.32\times 10^{-3}x\!-\! 1.32\times 10^{-3}\hat{x} \!+\! 8.93\times 10^{-5}x^2+1.28\times 10^{-4}x \hat{x}+8.93\times 10^{-5}\hat{x}^2, \text{ if } q = \bar{q}_0;\\
    &\!-\! 1.04\times 10^{-4}\!-\! 1.86\times 10^{-3}x\!-\! 1.86\times 10^{-3}\hat{x}\!+\! 1.50\times 10^{-4}x^2+1.30\times 10^{-4}x \hat{x}+1.50\times 10^{-4}\hat{x}^2, \text{ if } q = q_1;\\
    &\!-\!1.04\times 10^{-4}\!-\!1.86\times 10^{-3}x\!-\!1.86\times 10^{-3}\hat{x} \!+\! 1.50\times 10^{-4}x^2+1.30\times 10^{-4}x \hat{x}+1.50\times 10^{-4}\hat{x}^2, \text{ if } q = q_2;\\
    &\!-\!1.13\times 10^{-4}\!-\!1.88\times 10^{-3}x \!-\!1.88\times 10^{-3}\hat{x}\!+\!1.51\times 10^{-4}x^2+1.31\times 10^{-4}x \hat{x}+1.51\times 10^{-4}\hat{x}^2, \text{ if } q = q_{\text{trap}};\\
    &\!-\!1.04\times 10^{-4}\!-\!1.86\times 10^{-3}x\!-\! 1.86\times 10^{-3}\hat{x}\!+\! 1.50\times 10^{-4}x^2+1.30\times 10^{-4}x \hat{x}+1.50\times 10^{-4}\hat{x}^2, \text{ if } q = \bar{F}.\\
	\end{aligned}\right.
\end{align*}
which can be computed by solving a series of linear programming problems considering those constraints in~\eqref{rw21}-\eqref{rw23} since the system in Figure~\ref{runningexample} has finite state and input sets.
\end{example}

\subsubsection{CEGIS-based Approach}
Similar to Section~\ref{c-based_method1}, to compute a $\mathcal{V}$-HBC using SMT solvers under the CEGIS framework, we first reformulate conditions~\eqref{rw21}-\eqref{rw23} as in~\eqref{finrw21}-\eqref{finrw23}, respectively:
\begin{align}
    &\overline{\Phi}_1:= \bigwedge_{q\in Q_{\text{init}},(x,\hat{x})\in \mathcal{G}(\bar{q}_0,q)}\!\!\!\!\! \!\!\!\!\!\!\!\!\!\!\mathcal{V}_{\Delta(q)}(x_j,\hat{x}_j)\leq 0;\label{finrw21}\\
    &\overline{\Phi}_2:=\bigwedge_{q\in \bar{Q}\setminus \bar{F},(x,\hat{x})\in \tilde{P}(\mathcal{R})\setminus \mathcal{R}} \! \!\!\!\!\!\!\!\!\!\!\mathcal{V}_{\Delta(q)}(x,\hat{x}) > 0;\label{finrw22}\\
    &\overline{\Phi}_3:=\!\!\!\!\!\!\bigwedge_{\substack{q\in \bar{Q}\setminus \bar{F},\\(x,\hat{x})\in \overline{\mathcal{R}} }} \bigvee_{\substack{q'\in \text{Nxt}(q), \\ (x,\hat{x}, u,\hat{u}) \in \bar{X}(q,q')}}\!\!\!\!\!\!\!\!\!\!\!\!\!\mathcal{V}_{\Delta(q')}(f(x,u),f(\hat{x},\hat{u})))< \mathcal{V}_{\Delta(q)}(x,\hat{x}),\label{finrw23}
\end{align}
where $q' := \bar{\tau}(q,\bar{L}((f(x,u),f(\hat{x},\hat{u}))))$, sets $Q_{\text{init}}$, $\mathcal{G}(\bar{q}_0,q)$, and $\text{Nxt}(q)$ are as in~\eqref{set1}, \eqref{setg}, and  \eqref{next_state}, respectively.
Then, we collect $J$ points from $\mathcal{R}$ to obtain the set $\mathcal{D}'_x:=\{(x_1,\hat{x}_1),\ldots,(x_J,\hat{x}_J)\}$, and compute function $\mathcal{V}(x,\hat{x}, q)$ as in~\eqref{eq:V} such that $\overline{\Phi}_1 \wedge \overline{\Phi}_2 \wedge \overline{\Phi}_3$ is satisfied.
If no such $\mathcal{V}(x,\hat{x}, q)$ can be found, one can conclude that there is no function $\mathcal{V}(x,\hat{x}, q)$ with the selected template satisfying~\eqref{finrw21}-\eqref{finrw23} simultaneously. 
Otherwise, similar to the CEGIS-based approach for computing $\mathcal{B}(x,\hat{x},q)$ as proposed in Section~\ref{c-based_method1}, we encode $\neg (\overline{\Phi}_1 \wedge \overline{\Phi}_2 \wedge \overline{\Phi}_3)$ as an SMT query and search for counterexample $(x'_c,\hat{x}'_c)$ such that $\neg (\overline{\Phi}_1 \wedge \overline{\Phi}_2 \wedge \overline{\Phi}_3)$ is true.
If no such counterexamples can be found, one can conclude that $\mathcal{V}(x,\hat{x}, q)$ obtained based on the set $\mathcal{D}'_x$ is a $\mathcal{V}$-HBC as introduced in Theorem~\ref{veri_Lack_Diag}.
Otherwise, one should add the counterexample to the set $\mathcal{D}'_x$ and repeat the process.
Here, we summarize the CEGIS-based computation of $\mathcal{V}$-HBC in Algorithm~\ref{alg:3}.

\IncMargin{0.5em}
\begin{algorithm2e}[ht!]
	\DontPrintSemicolon
	\Indm 
	\KwIn{A dt-CS $\Sigma:=(X, X_0,X_F,U,f,Y,h)$, a ($\delta$,$K$)-DFA $\mathcal{A}_{(\delta,K)}\!=\! (\bar{Q},\bar{q}_0, \bar{\Pi},\bar{\tau},\bar{F})$ with a labeling function $\bar{L}: \mathcal{R}\rightarrow \Pi$ as in Definition~\ref{dKDFA}, and a template of function $\mathcal{V}(x,\hat{x},q)$ as in~\eqref{eq:V}, and $\mathsf{i}_{\text{max}}$ as the maximal number of iteration.}
	\KwOut{A $\mathcal{V}$-HBC $\mathcal{V}(x,\hat{x},q)$ if the computation terminates successfully.}
	\Indp
    $\mathsf{i}=0$, build the set $\mathcal{D}'_x:=\{(x_1,\hat{x}_1),\ldots,(x_J,\hat{x}_J)\}$.\\
    \While{1}
	{
        Compute the coefficients of $\mathcal{V}(x,\hat{x},q)$ such that~\eqref{finrw21}-\eqref{finrw23} holds for all $(x,\hat{x})\in\mathcal{D}'_x$.\\
        \eIf{$\mathcal{V}(x,\hat{x},q)$ cannot be found}{
            The $\mathcal{V}$-HBC with the given template doesn't exist.\\
            Terminate unsuccessfully.
        }
        {
            Compute counterexamples $(x'_c,\hat{x}'_c)$ satisfying $\neg (\overline{\Phi}_1 \wedge \overline{\Phi}_2 \wedge \overline{\Phi}_3)$ for the obtained $\mathcal{V}(x,\hat{x},q)$.\\
            \eIf{$(x'_c,\hat{x}'_c)$ cannot be found}{
            $\mathcal{V}(x,\hat{x},q)$ is a $\mathcal{V}$-HBC.\\
            Terminate successfully.
            }
            {
            $\mathcal{D}'_x:=\mathcal{D}'_x \cup \{ (x'_c,\hat{x}'_c)\}$.\\
            $\mathsf{i}=\mathsf{i}+1$.
            }
        }
        \If{$\mathsf{i} > \mathsf{i}_{\text{max}}$}{
        Terminate unsuccessfully.
        }
    }
    \caption{CEGIS-based computation of $\mathcal{V}$-HBC as in Theorem~\ref{veri_Lack_Diag}.} 
	\label{alg:3}
\end{algorithm2e}
\DecMargin{0.5em}

\subsubsection{SOS-based Approach}
To compute a $\mathcal{V}$-HBC over the product system $\Sigma_{\text{aug}} \otimes \mathcal{A}_{(\delta,K)}$, apart from Assumption~\ref{ass1}, the following assumption is required.
\begin{assumption}\label{ass2} 
	Consider a dt-CS $\Sigma:=(X, X_0,X_F, $ $U,f, Y,h)$ in Definition~\ref{def:sys1} such that Assumption~\ref{ass1} holds.
	We further assume that $U$ is in the form of polytope, i.e.	 $U \!=\! \{ u \in\mathbb{R}^{m}~|~ A_u u\!-\! b_u\!\geq\! 0 \}$, with $A_u \!\in\! \mathbb{R}^{\textsf{m}\!\times\! m}$, $b_u \!\in \! \mathbb{R}^{\mathsf{m}\!\times\! 1}$, and $\mathsf{m}\!\in\! \mathbb{N}_{>0}$, where the inequalities are component-wise.  
\end{assumption}

  Moreover, the following definition is also required for introducing the SOS-based method to compute a $\mathcal{V}$-HBC.
\begin{assumption}\label{def:function2}
    Consider a dt-CS $\Sigma:=(X, X_0,X_F,$ $ U,f, Y,h)$ as in Definition~\ref{def:sys1}.
    We assume that we are given vectors of polynomial functions $g_3(x,\hat{x})$ and $\bar{g}_1(x,\hat{x})$ of appropriate sizes such that:
    \begin{align}
        &(\tilde{P}(\mathcal{R})\setminus \mathcal{R}) \subseteq \!\!\{ (x,\hat{x})\in\ \mathbb{R}^{2n} ~|~ g_3(x,\hat{x}) \geq 0 \},\label{help1}\\
        &\overline{\mathcal{R}} \subseteq \{ (x,\hat{x})\in\ \mathbb{R}^{2n}  ~|~ \bar{g}_1(x,\hat{x})\geq 0 \},\label{eq_bg1}
\end{align}
in which all inequalities are component-wise. 
\end{assumption}

In the remainder of the paper, we omit $(x,\hat{x})$ in~\eqref{help1} and~\eqref{eq_bg1} when they are clear from the context.
Now, we are ready to propose the SOS-based scheme to compute a $\mathcal{V}$-HBC in Theorem~\ref{veri_Lack_Diag}.
	\begin{lemma}\label{lemma2}
		Consider a dt-CS $\Sigma:=(X, X_0,X_F, U,f, Y,h)$  in Definition~\ref{def:sys1} such that Assumption~\ref{ass2} holds. Consider constants $\delta \in\mathbb{R}_{>0}$, $K \in\mathbb{N}$, the ($\delta$,$K$)-DFA $\mathcal{A}_{(\delta,K)}=(\bar{Q},\bar{q}_0, \bar{\Pi},\bar{\tau},\bar{F})$, and functions $g_0(q)$, $\bar{g}_1$, and $g_3$ as in~\eqref{eq_bg}, \eqref{eq_bg1}, and~\eqref{help1}, respectively.
		Suppose that there exists a function $\mathcal{V}(x,\hat{x},q)$ as in~\eqref{eq:V} such that the following expressions are SOS polynomials:
        \begin{align}
			&\forall q \in Q_{\text{init}}, \quad -\mathcal{V}_{\Delta(q)}(x,\hat{x})-\beta_0^\top(q)g_0(q);\label{SOS21}\\
			& \forall q\in \bar{Q}\setminus \bar{F},  \quad \mathcal{V}(x,\hat{x},q)-\beta_1^\top(q)g_3-\varepsilon_1;\label{SOS22}\\
			& \forall q \!\in\! \bar{Q}\!\setminus\! \bar{F}, \forall q' \!\in\! \text{Nxt}(q), \,  -\mathcal{V}_{\Delta(q')}(f(x,u),f(\hat{x},\hat{u}))\!+\!\mathcal{V}_{\Delta(q)}\!(x,\hat{x})\nonumber\\
   &\quad \quad \quad \quad \quad \quad \quad \quad \quad \quad \quad \quad -\!\sum_{i=1}^{m}\big(u_i\!-\!\rho^{(q,q')}_i(x,\hat{x})\big)\!-\sum_{i=1}^{m}\big(\hat{u}_i \!-\! \hat{\rho}^{(q,q')}_i(x,\hat{x})\big)\!-\!\beta_2^\top(q) \bar{g}_1-\varepsilon_2;\label{SOS23}\\
			& \forall q\in \bar{Q}\setminus \bar{F}, \forall q' \in \text{Nxt}(q),\forall j \!\in\! [1,\mathsf{m}], A_{r}(j)\rho^{(q,q')}(x,\hat{x}) \!- \! b_r(j)  \!- \! \beta_{3,j}^\top(q,q') \bar{g}_1;\label{SOS24}\\
			& \forall q\in \bar{Q}\setminus \bar{F}, \forall q' \in \text{Nxt}(q), \forall  e\!\in\![1,\mathsf{m}],  A_{r}(e)\hat{\rho}^{(q,q')}(x,\hat{x})\!-\!b_r(j)\!-\!\beta_{4,e}^\top (q,q')\bar{g}_1,\label{SOS25}
		\end{align} 
        $\!\!$in which $\varepsilon_1,\varepsilon_2>0$, $\text{Nxt}(q)$ is defined as in~\eqref{next_state}; 
        $A_r$ and $b_r$ are defined as in Assumption~\ref{ass2};
        $\forall q\in \bar{Q}\setminus \bar{F}, \forall q' \in \text{Nxt}(q)$,
        \begin{align*}
            \rho^{(q,q')}(x,\hat{x}):=[\rho^{(q,q')}_1(x,\hat{x});\ldots;\rho^{(q,q')}_m(x,\hat{x})],   \hat{\rho}^{(q,q')}(x,\hat{x}):=[\hat{\rho}^{(q,q')}_1(x,\hat{x});\ldots;\hat{\rho}^{(q,q')}_m(x,\hat{x})],
        \end{align*}
        are vectors of polynomials;
        for any $q,q'\in\bar{Q}$, $\beta_0(q)$, $\beta_1(q)$, $\beta_{3,j}(q,q')$ and $\beta_{4,e}(q,q')$, with $j \!\in\! [1,\mathsf{m}]$ and $e\!\in\![1,\mathsf{m}]$, (resp. $\beta_2(q)$) are vectors of SOS polynomials in variables $x$ and $\hat{x}$ (resp. $x$, $\hat{x}$, $u$, and $\hat{u}$) of appropriate sizes.
		Then, function $\mathcal{V}(x,\hat{x},q)$ satisfies~\eqref{cd2_1}-\eqref{cd2_3}.  
	\end{lemma}

The proof of Lemma~\ref{lemma2} is provided in the Appendix.
So far, we have discussed how to compute $\mathcal{B}$-HBC (resp. $\mathcal{V}$-HBC) to verify (the lack of) $\delta$-approximate $K$-diagnosability.
In the next section, a design of a ($\delta$,$K$)-diagnoser as in Definition~\ref{def:diagnoser} will be proposed.

\section{Designing Diagnosers for dt-CS}\label{sec:Diagnoser}
In this section, we discuss how to design a diagnoser to detect the occurrence of  fault if the dt-CS is shown to be $\delta$-approximate $K$-diagnosable by leveraging the results in Sections~\ref{sec3} and~\ref{sec4}.
First, we propose Theorem~\ref{exist_diagnoser}, which shows that the existence of a ($\delta$,$K$)-diagnoser as in Definition~\ref{def:diagnoser} is actually equivalent to the system being $\delta$-approximate $K$-diagnosable.
\begin{resp}
\begin{theorem}\label{exist_diagnoser}
	Consider a dt-CS $\Sigma:=(X, X_0,X_F,$ $ U,f,Y,h)$.
	Given constants $\delta\in\mathbb{R}_{\geq 0}$ and $K\in\mathbb{N}$, there exists a ($\delta$,$K$)-diagnoser as in Definition~\ref{def:diagnoser} if and only if $\Sigma$ is $\delta$-approximate $K$-diagnosable.
\end{theorem}
\end{resp}

The proof of Theorem~\ref{exist_diagnoser} is provided in the Appendix.
Next, we define a sequence of sets
\begin{align}
	M(k)\!:=\! \big( \{ x\!\in\! X \!:\! \exists u\!\in\! U, x'\!\in\! M(k-1), x\! = \!f(x',u)\} \cap\{x \in X \!: \!||h(x) \!-\! \mathbf{y}^{\delta}_{x_0,\nu}(k) ||\!\leq\! \delta\}\big)\!\setminus\! X_F, k\in\mathbb{N}_{>0},\label{Mk}
\end{align}
which is initialized as 
\begin{equation}
M(0):=\{x \in X_0: ||h(x) - \mathbf{y}^{\delta}_{x_0,\nu}(0) ||\leq \delta\}\setminus X_F.
\end{equation}
With this definition, we are ready to propose the design of a ($\delta$,$K$)-diagnoser in the following results.
\begin{resp}
	\begin{theorem}\label{observer}
		Consider a dt-CS $\Sigma:=(X, X_0,X_F,$ $U,f,Y,h)$, which is $\delta$-approximate $K$-diagnosable.
	Given a state run $\mathbf{x}_{x_0,\nu}$, the corresponding imprecise observation $\mathbf{y}^{\delta}_{x_0,\nu}$ as in~\eqref{imprecise}, one has
  \begin{enumerate}
      \item $\exists k'\in \mathbb{N}, \mathbf{x}_{x_0,\nu}(k')\in X_F \Rightarrow \exists i\in [0,k'+K], M(i)=\emptyset$,
      \item $\exists \mathsf{i}\in \mathbb{N}$, with $\mathsf{i}\geq K$, such that $M(\mathsf{i})=\emptyset$, and $M(j)\neq \emptyset$, $\forall j\in[0,\mathsf{i}-1]$ $\Rightarrow$
            $\exists k'\in [\mathsf{i-K}, \mathsf{i}]$, $\mathbf{x}_{x_0,\nu}(k')\in X_F$,
  \end{enumerate}
  with the set $M(k)$, $k\in\mathbb{N}$, being defined as in~\eqref{Mk}.
	\end{theorem}
\end{resp}
The proof of Theorem~\ref{observer} is provided in the Appendix.
Intuitively, \text{(i)} in Theorem~\ref{observer} indicates that if $\mathbf{x}_{x_0,\nu}$ enters the faulty set $X_F$ at time step $k'$, then the set $M(i)$ as in~\eqref{Mk} would become empty at some time step $i\in [0,k'+K]$.
Additionally, \text{(ii)} implies that if the set $M(\mathsf{i})$ becomes an empty set for the first time at time step $\mathsf{i}$, then one can conclude that $\mathbf{x}_{x_0,\nu}$ has entered the faulty set $X_F$ in the time window $[\mathsf{i}-K, \mathsf{i}]$.
Therefore, one can construct a ($\delta$,$K$)-diagnoser as introduced in Definition~\ref{def:diagnoser} based on Theorem~\ref{observer} with running mechanism as described in Algorithm~\ref{alg:implementation}.
Note that the termination of the algorithm indicates an occurrence of a fault.
Additionally, the set $M'(k)$ in step 4 of Algorithm~\ref{alg:implementation} can be computed via existing toolboxes such as \texttt{CORA}~\cite{Althoff2015a}, \texttt{MPT3}~\cite{MPT3}, and \texttt{JuliaReach}~\cite{Bogomolov2019Julia}.
In case the set $M'(k)$ cannot be computed precisely, one can compute a (tight) inner approximation of $M'(k)$ so that all faults can be detected.
Before proceeding to the case study, we revisit the running example to provide more intuition for the running mechanism of the ($\delta$,$K$)-diagnoser as described in Algorithm~\ref{alg:implementation}.

\IncMargin{0.5em}
\begin{algorithm2e}[ht!]
	\DontPrintSemicolon
	\Indm 
	\KwIn{A dt-CS $\Sigma:=(X, X_0,X_F,U,f,Y,h)$ and an imprecise observation $\mathbf{y}^{\delta}_{x_0,\nu}$ as in~\eqref{imprecise}.}
	\KwOut{A fault occurred within the time window $[\text{max}(k-K,0),k]$.}
	\Indp
    $k=0$, compute $M(0)$ based on $\mathbf{y}^{\delta}_{x_0,\nu}(0)$.\\
    \While{$1$}
	{
        $k=k+1$.\\
        Compute $M'(k) := \{ x\in X : \exists u\in U, \exists x'\in M(k-1), x = f(x',u)\}$.\\
        Obtain $\mathbf{y}^{\delta}_{x_0,\nu}(k)$ and compute $M''(k) := \{x \in X : ||h(x) - \mathbf{y}^{\delta}_{x_0,\nu}(k) ||\leq \delta\}$.\\
        Compute $M(K):= (M'(k)\cap M''(k))\setminus X_F$.\\
        \If{$M(k)= \emptyset$}{
            Occurrence of a fault within the time window $[\text{max}(k-K,0),k]$ is detected.\\
            Computation terminates.
        }
    }
	\caption{Running mechanism of the ($\delta$,$K$)-diagnoser constructed based on Theorem~\ref{observer}.}
	\label{alg:implementation}
\end{algorithm2e}
\DecMargin{0.5em}

\addtocounter{example}{-1}
\begin{example}[continued]
    Since we found a $\mathcal{B}$-HBC as in~\eqref{eq:B1} for the running example, then it is $\delta$-approximate $K$-diagnosable with $\delta=1$ and $K=3$ according to Theorem~\ref{thm:1} and Theorem~\ref{veri_Diag}. 
    Here, we simulate the system with two input runs $\nu = 1,2,1,2$ and $\nu'= 2,1,2,1$, and we obtain the imprecise output runs 
    $\mathbf{y}_{x_0,\nu}=0,2.2,3.2,5.2,7.2$ and $\mathbf{y}_{x_0,\nu'}=0,1.2,3.2,5.2,9$, respectively.
    
    According to the dynamics of the running example as in Figure~\ref{runningexample}, one can verify that $\mathbf{x}_{x_0,\nu}(1)\in X_F$ and $\mathbf{x}_{x_0,\nu'}(k)\notin X_F$ for all $i\in[0,4]$. 
    Therefore, by applying a ($\delta$,$K$)-diagnoser $D$ running as in Algorithm~\ref{alg:implementation}, we obtain the sequences of the set $M(k)$ as in~\eqref{Mk} associated with $\mathbf{y}_{x_0,\nu}$ and $\mathbf{y}_{x_0,\nu'}$ as $\{0\}\{2.2\}\{4.2\}\{6.2\}\{\emptyset\}$ and $\{0\}\{2.2\}\{4.2\}\{6.2\}\{9\}$, respectively. 
    Therefore, one gets $D(\mathbf{y}_{x_0,\nu})=1$ and $D(\mathbf{y}_{x_0,\nu'})=0$, with function $D$ being defined as in~\eqref{Dfunction}, showing that the diagnoser running as in Algorithm~\ref{alg:implementation} fulfills conditions (C1) and (C2) as in Definition~\ref{def:diagnoser}. 
\end{example}

\section{Case Study}\label{sec:case}
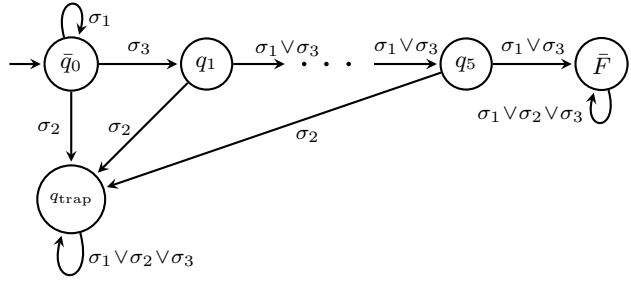
\begin{figure} 
	\centering
	\centering
	\begin{tikzpicture}[->,>=stealth,shorten >=1pt,auto,node distance=1.8cm,
		thick,base node/.style={circle,draw,minimum size=0.5mm, font=\small},]
		\node[initial,initial text={}, base node] (q0) at (0,1) {$\bar{q}_0$};
		\node[state, base node](q1) [right of= q0] {$q_1$};
        \node[font=\Huge] (q2) at ($(q1.east)!.5!(q3.west)$) {\ldots};
		\node[state, base node](q3) [right of= q2] {$q_5$};
		\node[state, base node](F) [right of= q3] {$\bar{F}$};
		
		\node[state, base node, font=\tiny](tr) [below of= q0] {$q_{\text{trap}}$};

		\path (q0) edge node {\fontsize{8}{1}$\sigma_3$} (q1)
		(q1) edge node [right,xshift=-0.2cm, yshift=0.2cm]{\fontsize{8}{1}$\sigma_1  \! \vee  \! \sigma_3$} (q2)
		(q2) edge node {\fontsize{8}{1}$\sigma_1  \! \vee  \! \sigma_3$} (q3)
		(q3) edge node {\fontsize{8}{1}$\sigma_1  \! \vee  \! \sigma_3$} (F);
		
		\path (q0) edge node[left] {\fontsize{8}{1}$\sigma_2$} (tr)
		(q1) edge node[left] {\fontsize{8}{1}$\sigma_2$} (tr)
		(q3) edge node[right=0.4,pos=0.55] {\fontsize{8}{1}$\sigma_2$} (tr);
		
		\path[] (q0) edge [loop above] node [right,xshift=0.1cm, yshift=-0.2cm] {\fontsize{8}{1}$\sigma_1$} (q0);
		\path[] (tr) edge [loop below] node [right,xshift=0.1cm, yshift=0.2cm] {\fontsize{8}{1}$\sigma_1 \! \vee  \! \sigma_2  \! \vee  \! \sigma_3$} (tr);
		\path[] (F) edge [loop below] node [left=0.1,yshift=0.1cm] {\fontsize{8}{1}$\sigma_1  \! \vee  \! \sigma_2 \! \vee \! \sigma_3$} (F);
		
	\end{tikzpicture}
		\caption{The $(0.5,5)$-DFA for the case study of a two-room temperature model in~\eqref{eq:casestudy}.}\label{DFA_temp}
\end{figure}

To demonstrate the effectiveness of the proposed results, we apply them to a case study of a two-room temperature model borrowed from \cite{meyer2017compositional} with dynamics:
\begin{align}\label{eq:casestudy}
	\!\!\!\!\Sigma\!:\!\left\{
	\begin{array}{rl}
		x_1(k+1)=& \!(1\!-\!2\alpha\!-\!\alpha_e\!-\!\alpha_h u_1(k))x_1(k)\!+\alpha x_2(k)+\alpha_h T_h u_1(k)+ \alpha_e T_e,\\
		x_2(k+1)= &\!(1\!-\!2\alpha\!-\!\alpha_e\!-\!\alpha_h u_2(k))x_2(k)+\alpha x_1(k)+\alpha_h T_h u_2(k)+ \alpha_e T_e,\\
		y(k)=& \![x_1(k);x_2(k)], \quad \quad \quad  \ \,k\in\mathbb{N},
	\end{array}
	\right.
\end{align}
with state set $X:={[15,30]}^2$, initial state set $X_0:={[19.5,20.5]}^2$, faulty state set $X_F:=[24,26]^2$, and input state set $U:=[0,1]^2$, 
in which $\alpha=0.01$, $\alpha_e=0.04$, and $\alpha_h=0.145$ are heat exchange coefficients; $T_e=10^\circ C$ is the external temperature; and $T_h=50^\circ C$ is the heater temperature.
Here, we are interested in the $\delta$-approximate $K$-diagnosability with 1) $\delta =0.5$ and $K =5$ and 2) $\delta =0.5$ and $K =3$.
The ($\delta$,$K$)-DFA as in Definition~\ref{dKDFA} associated with the $0.5$-approximate $5$-diagnosability is demonstrated in Figure~\ref{DFA_temp}, with alphabet $\Pi = \{\sigma_1,\sigma_2,\sigma_3 \}$ and the labeling function $\bar{L}$ defined as 
$\bar{L}^{-1}(\sigma_1):= \{(x,\hat{x}):||h(x)-h(\hat{x})||\leq 0.5, x,\hat{x}\notin X_F\}$,
$\bar{L}^{-1}(\sigma_2):= \{(x,\hat{x}):||h(x)-h(\hat{x})||> 0.5\} \cup \{(x,\hat{x}):||h(x)-h(\hat{x})||\leq 0.5, \hat{x}\in X_F\}$,
and $\bar{L}^{-1}(\sigma_3):= \{(x,\hat{x}):||h(x)-h(\hat{x})||\leq 0.5, x \in X_F, \hat{x}\notin X_F\}$.
The ($\delta$,$K$)-DFA associated with the $0.5$-approximate $3$-diagnosability can be constructed in a similar way and we omit it here for the lack of space.

{\bf Computation of the $\mathcal{B}$-HBC}:
Having the ($\delta$,$K$)-DFA as in Figure~\ref{DFA_temp} associated with the $0.5$-approximate $5$-diagnosability, we construct the product system $\Sigma_{\text{aug}} \otimes \mathcal{A}_{(\delta,K)}$ as in Definition~\ref{def:product_dtCS}, and we consider a template of $\mathcal{B}$-HBC as in~\eqref{eq:2}, with $\mathcal{B}_i$ being polynomial functions of order 2, for all $i\in [0, 7]$.
By running Algorithm~\ref{alg:CEGIS}, we find a $\mathcal{B}$-HBC (see~\eqref{Bcase} in the Appendix),
indicating that the system in~\eqref{eq:casestudy} is $0.5$-approximate $5$-diagnosable.
Meanwhile, one can compute the $\mathcal{B}$-HBC (see~\eqref{Bcase-2} in the Appendix) by SOS-based approach, which satisfies~\eqref{soscond11}-~\eqref{soscond13} with 4-order SOS polynomial multipliers.

{\bf Computation of the $\mathcal{V}$-HBC}:
Considering the $0.5$-approximate $3$-diagnosability property, we construct the ($\delta$,$K$)-DFA and the product system $\Sigma_{\text{aug}} \otimes \mathcal{A}_{(\delta,K)}$ as in Definition~\ref{dKDFA} and Definition~\ref{def:product_dtCS}, respectively.
Additionally, we consider a template of $\mathcal{V}$-HBC as in~\eqref{eq:V}, with $\mathcal{V}_i$ being polynomial functions of order 2, for all $i\in [0,5]$.
By applying Algorithm~\ref{alg:3}, we find a $\mathcal{V}$-HBC (see~\eqref{Bcase-V-1} in the Appendix),
indicating that the system in~\eqref{eq:casestudy} is \emph{not} $0.5$-approximate $3$-diagnosable. 
Moreover, one can compute the $\mathcal{V}$-HBC by SOS-based approach (see~\eqref{Bcase-V-2} in the Appendix), which satisfies~\eqref{SOS21}-\eqref{SOS25} with 4-order SOS polynomial multipliers.

\begin{figure}[t]
	\centering
	\includegraphics[width = 240pt]{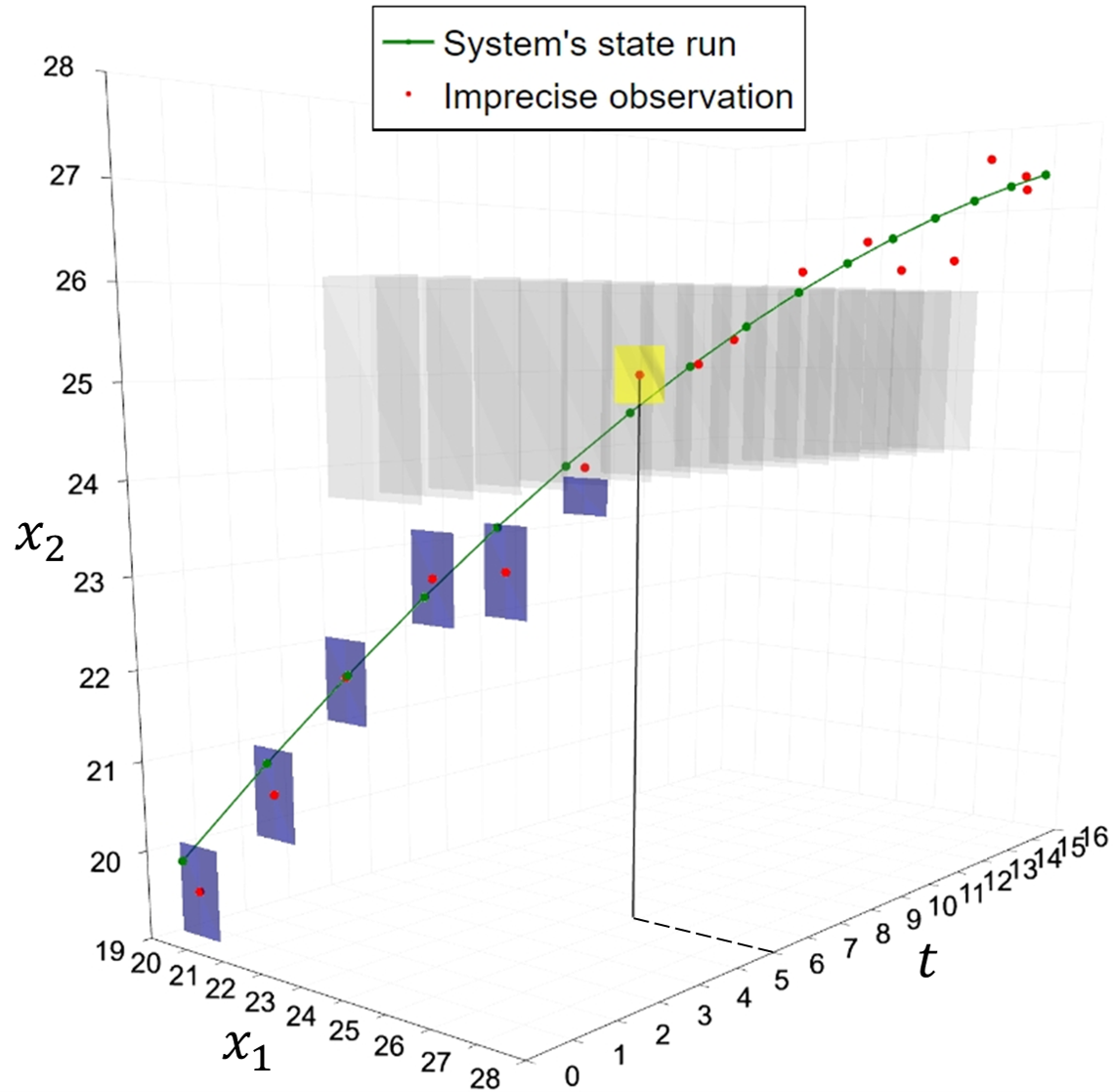}
	\caption{Simulation of two-room temperature model starting from initial region $[19.5,20.5]^2$, where green dots are the actual state of the system and red dots are imprecise observations that are obtained. Faulty region is denoted by gray rectangles at each time step. Sets $M(k)$ are illustrated by blue rectangles and the empty $M(k)$ (at time step $k=7$) is shown by a yellow cube. }
	\label{fig:case1}
\end{figure}

{\bf Demonstration of the diagnoser considering the $0.5$-approximate $5$-diagnosability}:
To demonstrate the diagnoser with running mechanism as in Algorithm~\ref{alg:implementation}, we simulate the system with control policy\footnote{Note that the controller here is only used for demonstrating the effectiveness of the diagnoser. 
In the context of verification of diagnosability property, we try to diagnose a fault based on the imprecise observation regardless of the control inputs applied to the system.} $u(k)=[0.5;0.5], \forall k \in \N_{\geq 0}$.
The state run and the imprecise observation of the system are depicted in Figure~\ref{fig:case1}.
We can see that the system visited the faulty state set $X_F$ at time step $k=5$.
The evolution of the under approximation of $M(k)$, which is computed using \texttt{JuliaReach}~\cite{Bogomolov2019Julia}, is shown in {Figure~\ref{fig:case1}} by blue rectangles at each time steps.
At time step $k=6$, the set $M(k)$ becomes empty (highlighted by a yellow cube), indicating that a fault has happened within the time interval $[1,5]$ according to Theorem~\ref{observer}.
In other words, the fault is diagnosed within $5$ step after its appearance, which meets the requirement for the $0.5$-approximate $5$-diagnosability property.

\section{Conclusion}\label{sec:conclusion}
In this paper, we developed an abstraction-free, automata-based framework for verifying $\delta$-approximate $K$-diagnosability using hybrid barrier certificates (HBC) over discrete-time control systems with both continuous and discrete states and input sets. 
Initially, we designed a ($\delta$,$K$)-DFA that tracks the number of time steps following fault occurrences. 
This allows the verification of (the lack of) $\delta$-approximate $K$-diagnosability to be recasted as a safety verification problem over a product system that integrates the ($\delta$,$K$)-DFA with an augmented version of the discrete-time control system.
We further show that the safety verification problem over the product system can be addressed by computing appropriate HBCs.
Accordingly, we introduced two methodologies for computing HBCs, utilizing sum-of-squares programming and counter-example-guided inductive synthesis. 
Moreover, we propose the construction of a diagnoser to identify faults in systems that are $\delta$-approximate $K$-diagnosable. 
Our future work aims to extend the current result to fault models described by   language patterns \cite{lefebvre2022diagnosability,ma2024verification} or temporal logics   \cite{7167720,dong2023diagnosis}.

\bibliographystyle{elsarticle-num}   
\bibliography{sample-base} 

\begin{thebibliography}{10}
\expandafter\ifx\csname url\endcsname\relax
  \def\url#1{\texttt{#1}}\fi
\expandafter\ifx\csname urlprefix\endcsname\relax\def\urlprefix{URL }\fi
\expandafter\ifx\csname href\endcsname\relax
  \def\href#1#2{#2} \def\path#1{#1}\fi

\bibitem{sampath1995diagnosability}
M.~Sampath, R.~Sengupta, S.~Lafortune, K.~Sinnamohideen, D.~Teneketzis,
  Diagnosability of discrete-event systems, IEEE Transactions on Automatic
  control 40~(9) (1995) 1555--1575.

\bibitem{lu2007system}
T.-C. Lu, K.~W. Przytula, System diagnosability analysis using p-slop map., in:
  FLAIRS, 2007, pp. 116--121.

\bibitem{basile2017diagnosability}
F.~Basile, M.~Cabasino, C.~Seatzu, Diagnosability analysis of labeled time
  {P}etri net systems, IEEE Transactions on Automatic Control 62~(3) (2017)
  1384--1396.

\bibitem{lefebvre2007diagnosis}
D.~Lefebvre, C.~Delherm, Diagnosis of {DES} with {P}etri net models, IEEE
  Transactions on Automation Science and Engineering 4~(1) (2007) 114--118.

\bibitem{hu2021diagnosability}
Y.~Hu, Z.~Ma, Z.~Li, A.~Giua, Diagnosability enforcement in labeled {P}etri
  nets using supervisory control, Automatica 131 (2021) 109776.

\bibitem{ma2021marking}
Z.~Ma, X.~Yin, Z.~Li, Marking diagnosability verification in labeled {P}etri
  nets, Automatica 131 (2021) 109713.

\bibitem{pencole2022diagnosability}
Y.~Pencol{\'e}, A.~Subias, Diagnosability of event patterns in safe labeled
  time {P}etri nets: a model-checking approach, IEEE Transactions on Automation
  Science and Engineering 19~(2) (2022) 1151--1162.

\bibitem{ran2018codiagnosability}
N.~Ran, H.~Su, A.~Giua, C.~Seatzu, Codiagnosability analysis of bounded {P}etri
  nets, IEEE Transactions on Automatic Control 63~(4) (2018) 1192--1199.

\bibitem{yin2017decidability}
X.~Yin, S.~Lafortune, On the decidability and complexity of diagnosability for
  labeled {P}etri nets, IEEE Transactions on Automatic Control 62~(11) (2017)
  5931--5938.

\bibitem{thorsley2005diagnosability}
D.~Thorsley, D.~Teneketzis, Diagnosability of stochastic discrete-event
  systems, IEEE Transactions on Automatic Control 50~(4) (2005) 476--492.

\bibitem{chen2023probabilistic}
J.~Chen, A probabilistic test for a-diagnosability of stochastic discrete-event
  systems with guaranteed error bound, IEEE Control Systems Letters (2023).

\bibitem{yin2019robust}
X.~Yin, J.~Chen, Z.~Li, S.~Li, Robust fault diagnosis of stochastic discrete
  event systems, IEEE Transactions on Automatic Control 64~(10) (2019)
  4237--4244.

\bibitem{lin2017n}
F.~Lin, W.~Chen, L.~Han, B.~Shen, et~al., N-diagnosability for active on-line
  diagnosis in discrete event systems, Automatica 83 (2017) 220--225.

\bibitem{ma2023verification}
Z.~Ma, Y.~Tong, C.~Seatzu, Verification of pattern-pattern diagnosability in
  partially observed discrete event systems, IEEE Transactions on Automatic
  Control (2023).

\bibitem{takai2017generalized}
S.~Takai, R.~Kumar, A generalized framework for inference-based diagnosis of
  discrete event systems capturing both disjunctive and conjunctive
  decision-making, IEEE Transactions on Automatic Control 62~(6) (2017)
  2778--2793.

\bibitem{pola2017approximate}
G.~Pola, E.~De~Santis, M.~Di~Benedetto, Approximate diagnosis of metric
  systems, IEEE Control Systems Letters 2~(1) (2017) 115--120.

\bibitem{di2011verification}
M.~Di~Benedetto, S.~Di~Gennaro, A.~D'Innocenzo, Verification of hybrid automata
  diagnosability by abstraction, IEEE transactions on automatic control 56~(9)
  (2011) 2050--2061.

\bibitem{deng2016verification}
Y.~Deng, A.~D'Innocenzo, M.~Di~Benedetto, S.~Di~Gennaro, A.~Julius,
  Verification of hybrid automata diagnosability with measurement uncertainty,
  IEEE Transactions on Automatic Control 61~(4) (2016) 982--993.

\bibitem{bayoudh2008hybrid}
M.~Bayoudh, L.~Trav{\'e}-Massuyes, X.~Olive, Hybrid systems diagnosis by
  coupling continuous and discrete event techniques, IFAC Proceedings Volumes
  41~(2) (2008) 7265--7270.

\bibitem{dallal2013most}
E.~Dallal, S.~Lafortune, On most permissive observers in dynamic sensor
  activation problems, IEEE Transactions on Automatic Control 59~(4) (2013)
  966--981.

\bibitem{murali2023co}
V.~Murali, A.~Trivedi, M.~Zamani, Co-buchi barrier certificates for
  discrete-time dynamical systems, arXiv preprint arXiv:2311.07695 (2023).

\bibitem{zhong2024verification}
B.~Zhong, W.~Dong, Y.~X., Z.~M., Verification of diagnosability for
  cyber-physical systems via hybrid barrier certificates, in: Proceedings of
  the 8th IFAC Conference on Analysis and Design of Hybrid Systems (ADHS),
  IFAC, 2024, to appear.

\bibitem{yin2020approximate}
X.~Yin, M.~Zamani, S.~Liu, On approximate opacity of cyber-physical systems,
  IEEE Transactions on Automatic Control 66~(4) (2020) 1630--1645.

\bibitem{liu2020notion}
S.~Liu, X.~Yin, M.~Zamani, On a notion of approximate opacity for discrete-time
  stochastic control systems, in: 2020 American Control Conference (ACC), IEEE,
  2020, pp. 5413--5418.

\bibitem{pola2023approximate}
G.~Pola, E.~De~Santis, M.~Di~Benedetto, Approximate current state observability
  of discrete-time nonlinear systems under cyber-attacks, Nonlinear Analysis:
  Hybrid Systems 50 (2023) 101403.

\bibitem{kalat2021modular}
S.~Kalat, S.~Liu, M.~Zamani, Modular verification of opacity for interconnected
  control systems via barrier certificates, IEEE Control Systems Letters 6
  (2021) 890--895.

\bibitem{liu2020verification}
S.~Liu, M.~Zamani, Verification of approximate opacity via barrier
  certificates, IEEE Control Systems Letters 5~(4) (2020) 1369--1374.

\bibitem{zhao2024unified}
J.~Zhao, X.~Yin, S.~Li, A unified framework for verification of observational
  properties for partially-observed discrete-event systems, IEEE Transactions
  on Automatic Control (2024).

\bibitem{anand2021verification}
M.~Anand, V.~Murali, A.~Trivedi, M.~Zamani, Verification of hyperproperties for
  uncertain dynamical systems via barrier certificates, arXiv preprint
  arXiv:2105.05493 (2021).

\bibitem{Baier2008Principles}
C.~Baier, J.-P. Katoen, Principles of model checking, MIT press, 2008.

\bibitem{Lezama_2008_thesis}
A.~Solar-Lezama, Program Synthesis by Sketching, PhD thesis, University of
  California, Berkeley, 2008.

\bibitem{gurobi}
{Gurobi Optimization, LLC}, \href{https://www.gurobi.com}{{Gurobi Optimizer
  Reference Manual}} (2023).
\newline\urlprefix\url{https://www.gurobi.com}

\bibitem{ApS2019MOSEK}
{MOSEK ApS}, \href{http://docs.mosek.com/9.0/toolbox/index.html}{The MOSEK
  optimization toolbox for MATLAB manual. Version 9.3.6} (2019).
\newline\urlprefix\url{http://docs.mosek.com/9.0/toolbox/index.html}

\bibitem{JarvisWloszek2005Control}
Z.~Jarvis-Wloszek, R.~Feeley, W.~Tan, K.~Sun, A.~Packard, Control applications
  of sum of squares programming, in: Positive Polynomials in Control, Springer,
  2005, pp. 3--22.

\bibitem{moura_2008_z3}
L.~d. Moura, N.~Bj{\o}rner, Z3: An efficient {SMT} solver, in: International
  conference on Tools and Algorithms for the Construction and Analysis of
  Systems, 2008, pp. 337--340.

\bibitem{Althoff2015a}
M.~Althoff, An introduction to cora 2015, in: Proc. of the Workshop on Applied
  Verification for Continuous and Hybrid Systems, 2015, p. 120–151.
\newblock \href {https://doi.org/10.29007/zbkv} {\path{doi:10.29007/zbkv}}.

\bibitem{MPT3}
M.~Herceg, M.~Kvasnica, C.~Jones, M.~Morari, {Multi-Parametric Toolbox 3.0},
  in: Proc.~of the European Control Conference, Z\"urich, Switzerland, 2013,
  pp. 502--510, \url{http://control.ee.ethz.ch/~mpt}.

\bibitem{Bogomolov2019Julia}
S.~Bogomolov, M.~Forets, G.~Frehse, K.~Potomkin, C.~Schilling, Juliareach: a
  toolbox for set-based reachability, in: Proceedings of the 22nd ACM
  International Conference on Hybrid Systems: Computation and Control, HSCC
  '19, ACM, New York, USA, 2019, p. 39–44.

\bibitem{meyer2017compositional}
P.~Meyer, A.~Girard, E.~Witrant, Compositional abstraction and safety synthesis
  using overlapping symbolic models, IEEE Transactions on Automatic Control
  63~(6) (2017) 1835--1841.

\bibitem{lefebvre2022diagnosability}
D.~Lefebvre, C.~N. Hadjicostis, Diagnosability of fault patterns with labeled
  stochastic petri nets, Information Sciences 593 (2022) 341--363.

\bibitem{ma2024verification}
Z.~Ma, Y.~Tong, C.~Seatzu, Verification of pattern-pattern diagnosability in
  partially observed discrete event systems, IEEE Transactions on Automatic
  Control 69~(3) (2024) 2044--2051.

\bibitem{7167720}
J.~Chen, R.~Kumar, Fault detection of discrete-time stochastic systems subject
  to temporal logic correctness requirements, IEEE Transactions on Automation
  Science and Engineering 12~(4) (2015) 1369--1379.
\newblock \href {https://doi.org/10.1109/TASE.2015.2453193}
  {\path{doi:10.1109/TASE.2015.2453193}}.

\bibitem{dong2023diagnosis}
W.~Dong, S.~Li, X.~Yin, Diagnosis of time-sensitive failures in timed
  discrete-event systems with metric interval temporal logics, in: 62nd IEEE
  Conference on Decision and Control (CDC), IEEE, 2023, pp. 6827--6833.

\end{thebibliography}

\appendix

\renewcommand{\theequation}{A.\arabic{equation}}
\section{Proof of Statements}
\renewcommand{\theequation}{A.\arabic{equation}}
\renewcommand{\thetheorem}{A.\arabic{theorem}}
{\bf Proof for Theorem~\ref{thm:1}}
First, we prove the statement regarding \textit{if} by contradiction.
Suppose that $\nexists \tilde{\nu}, \tilde{\mathbf{x}}_{\tilde{x}_0,\tilde{\nu}} $ such that $\tilde{\mathbf{x}}_{\tilde{x}_0,\tilde{\nu}}(k)\in \mathcal{R}\times \bar{F}$ for some $k\in\mathbb{N}$, but the system is not $\delta$-approximate $K$-diagnosable.
    The lack of $\delta$-approximate $K$-diagnosability indicates that there exists $\mathbf{x}_{x_0,v}:= (x_0,\ldots,x_k,\ldots)$ with $x_0\in X_0$, $x_k \in X_F$ and $x_i\notin X_F$, $\forall i \in [0,k-1]$, such that there exists $ \hat{\mathbf{x}}_{\hat{x}_0,\hat{v}}:= (\hat{x}_0,\ldots,\hat{x}_k,\ldots)$ for which  $\hat{x}_0\in X_0$, $\hat{x}_i\notin X_F$, $\forall i\in[0,k+K]$, and $\max_{i\in [0,k+K]} \vert\vert  h(x_i)- h(\hat{x}_i)\vert \vert \leq \delta$.
Consider the state run $(\mathbf{x}_{x_0,v},\hat{\mathbf{x}}_{\hat{x}_0,\hat{v}})$ of the augmented system $\Sigma_{\text{aug}}$.
According to the definition of the ($\delta$,K)-DFA and the construction of the product system $\Sigma_{\text{aug}} \otimes \mathcal{A}_{(\delta,K)}$ as in Definition~\ref{def:product_dtCS}, considering the $\tilde{x}_{\tilde{x}_0,\tilde{\nu}}=\{(x(0),\hat{x}(0),q(0)),\ldots,(x(k),\hat{x}(k),$ $q(k)),\dots\}$, $k\in\mathbb{N}$, in which $x(k):=\mathbf{x}_{x_0,v}(k)$ and $\hat{x}(k):=\hat{\mathbf{x}}_{\hat{x}_0,\hat{v}}(k)$, one has $q(k+K)=\bar{F}$, which results in a contradiction to the fact that $\nexists \tilde{\nu}, \tilde{\mathbf{x}}_{\tilde{x}_0,\tilde{\nu}} $ such that $\tilde{\mathbf{x}}_{\tilde{x}_0,\tilde{\nu}}(k)\in \mathcal{R}\times \bar{F}$ for some $k\in\mathbb{N}$. Therefore, the statement regarding \textit{if} holds.

Then, we prove the statement regarding \textit{only if}.
Suppose the system is $\delta$-approximate $K$-diagnosable, meaning that for all $ \mathbf{x}_{x_0,v}:= (x_0,\ldots,x_k,\ldots)$ such that $x_0\in X_0$, $x_k \in X_F$, and $x_i\notin X_F$, $\forall i \in [0,k-1]$, and for all $ \hat{\mathbf{x}}_{\hat{x}_0,\hat{v}}:= (\hat{x}_0,\ldots,\hat{x}_k,\ldots)$ such that $\hat{x}_0\in X_0$, and $\hat{x}_i\notin X_F$, $\forall i\in[0,k+K]$, there exists $k'\in[0,k+K]$ such that $\vert\vert  h(x_{k'})- h(\hat{x}_{k'})\vert \vert>\delta$.
Considering the state run $\tilde{x}_{\tilde{x}_0,\tilde{\nu}}=\{(x(0),\hat{x}(0),q(0)),\ldots,(x(k),\hat{x}(k),q(k)),\dots\}$, $k\in\mathbb{N}$, of the product system $\Sigma_{\text{aug}} \otimes \mathcal{A}_{(\delta,K)}$ as in Definition~\ref{def:product_dtCS}, one has $q(k')=q_{\text{trap}}$. 
This immediately implies that $q(k)\neq \bar{F}$ for all $k\in\mathbb{N}$ so that the statement regarding \textit{only if} is true, which completes the proof.
$\hfill\blacksquare$

{\bf Proof for Theorem~\ref{veri_Diag}}
We prove Theorem~\ref{veri_Diag} by contradiction.
Suppose that there exists a function $\mathcal{B}$ such that~\eqref{cd1_1}-\eqref{cd1_3} hold, but the dt-CS $\Sigma$ is not $\delta$-approximate $K$-diagnosable.
In other words, there exists $ \mathbf{x}_{x_0,v}:= (x_0,\ldots,x_k,\ldots)$ with $x_0\in X_0$, $x_k \in X_F$ and $x_i\notin X_F$, $\forall i \in [0,k-1]$, such that there exists $ \hat{\mathbf{x}}_{\hat{x}_0,\hat{v}}:= (\hat{x}_0,\ldots,\hat{x}_k,\ldots)$ with $\hat{x}_0\in X_0$, $\hat{x}_i\notin X_F$, $\forall i\in[0,k+K]$, and $\max_{i\in [0,k+K]} \vert\vert  h(x_i)- h(\hat{x}_i)\vert \vert \leq\delta$.
Considering the definition of the ($\delta$,$K$)-DFA and the product system $\Sigma_{\text{aug}} \otimes \mathcal{A}_{(\delta,K)}$, one has $\tilde{\mathbf{x}}_{\tilde{x}_0,\tilde{\nu}}(k+K)\in \mathcal{R}\times \bar{F}$, with $\tilde{x}_0 := [x_0;\hat{x}_0]$, and $\tilde{\nu}:= ((\nu(0),\hat{\nu}(0)),\ldots,(\nu(k+K-1),\hat{\nu}(k+K-1)))$.
Note that if there exist $t\in [0,k+K]$ such that $\tilde{\mathbf{x}}_{\tilde{x}_0,\tilde{\nu}}(t)\in \mathcal{R}\times q_{\text{trap}}$, then one has $\tilde{\mathbf{x}}_{\tilde{x}_0,\tilde{\nu}}(t')\in \mathcal{R}\times q_{\text{trap}}$ for all $t'\geq t$.
Therefore, we only consider the case in which $\tilde{\mathbf{x}}_{\tilde{x}_0,\tilde{\nu}}(t)\notin \mathcal{R}\times q_{\text{trap}}$, $\forall t\in[0,k+K]$.
On one hand, according to~\eqref{cd1_1} and~\eqref{cd1_2}, one has $\mathcal{B}(\tilde{\mathbf{x}}_{\tilde{x}_0,\tilde{\nu}}(0))\leq 0$ and $\mathcal{B}(\tilde{\mathbf{x}}_{\tilde{x}_0,\tilde{\nu}}(k+K))> 0$.
On the other hand, one has $\mathcal{B}(\tilde{\mathbf{x}}_{\tilde{x}_0,\tilde{\nu}}(t'))\leq\mathcal{B}(\tilde{\mathbf{x}}_{\tilde{x}_0,\tilde{\nu}}(0))\leq 0$ for all $t''>0$ considering~\eqref{cd1_3}.
Note that this results in a contradiction that $\mathcal{B}(\tilde{\mathbf{x}}_{\tilde{x}_0,\tilde{\nu}}(k+K))> 0$.
As a results, for all $ \mathbf{x}_{x_0,v}:= (x_0,\ldots,x_k,\ldots)$ with $x_0\in X_0$ such that $x_k \in X_F$ and $x_i\notin X_F$, $\forall i \in [0,k-1]$, and for all $ \hat{\mathbf{x}}_{\hat{x}_0,\hat{v}}:= (\hat{x}_0,\ldots,\hat{x}_k,\ldots)$ with $\hat{x}_0\in X_0$ such that $\hat{x}_i\notin X_F$, $\forall i\in[0,k+K]$, one gets $\max_{i\in [0,k+K]} \vert\vert  h(x_i)- h(\hat{x}_i)\vert \vert > \delta$, indicating that it does not exist input run $\tilde{\nu}$ and state run $\tilde{\mathbf{x}}_{\tilde{x}_0,\tilde{\nu}} $ such that $\tilde{\mathbf{x}}_{\tilde{x}_0,\tilde{\nu}}(k)\in \mathcal{R}\times \bar{F}$ for some $k\in\mathbb{N}$, which completes the proof.
$\hfill\blacksquare$

{\bf Proof for Theorem~\ref{veri_Lack_Diag}}
According to~\eqref{cd2_1},  $\mathcal{V}(\tilde{x}_0)\leq 0$ holds for any $\tilde{x}_0\in\tilde{X}_0$.
Consider an input run $\tilde{\nu}$ for $\tilde{\mathbf{x}}_{\tilde{x}_0,\tilde{\nu}}$ such that~\eqref{cd2_3} holds.
Then, one can readily verify that there exists $\bar{t}\in \mathbb{N}_{>0}$ such that $\mathcal{V}(\tilde{\mathbf{x}}_{\tilde{x}_0,\tilde{\nu}}(\bar{t}))<\min_{\tilde{x}'\in \overline{\mathcal{R}}\times (\bar{Q}\setminus \bar{F})}\mathcal{V}(\tilde{x}')<\mathcal{V}(\tilde{\mathbf{x}}_{\tilde{x}_0,\tilde{\nu}}(\bar{t}-1))$, since~\eqref{cd2_3} indicates that $\mathcal{V}(\tilde{x})$ is strictly decreasing along $\tilde{\mathbf{x}}_{\tilde{x}_0,\tilde{\nu}}$ and the set $\overline{\mathcal{R}}\times (\bar{Q}\setminus \bar{F})$ is bounded.
In other words, there exists $\bar{t}\in \mathbb{N}_{>0}$ such that $\tilde{\mathbf{x}}_{\tilde{x}_0,\tilde{\nu}}(\bar{t}-1)\in \overline{\mathcal{R}}\times (\bar{Q}\setminus \bar{F})$ and $\tilde{\mathbf{x}}_{\tilde{x}_0,\tilde{\nu}}(\bar{t})\notin \overline{\mathcal{R}}\times (\bar{Q}\setminus \bar{F})$.
On the other hand, for all $\tilde{\mathbf{x}}_{\tilde{x}_0,\tilde{\nu}}(t')\in \overline{\mathcal{R}}\times (\bar{Q}\setminus \bar{F})$, with $t'\in\mathbb{N}$, one has $\tilde{\mathbf{x}}_{\tilde{x}_0,\tilde{\nu}}(t'+1)\notin (\tilde{P}(\mathcal{R})\setminus \mathcal{R}) \times (\bar{Q}\setminus\bar{F})$ considering~\eqref{cd2_2}, ~\eqref{cd2_3}, and the definition of the set $\tilde{P}(\mathcal{R})$.
Therefore, one has $\tilde{\mathbf{x}}_{\tilde{x}_0,\tilde{\nu}}(\bar{t})\in \mathcal{R}\times \bar{F}$.
$\hfill\blacksquare$

{\bf Proof for Lemma~\ref{lemma1}}
We prove Lemma~\ref{lemma1} by showing ~\eqref{soscond11}-\eqref{soscond13} are SOS polynomials implies~\eqref{cd1_1}-\eqref{cd1_3} holds, respectively.
\begin{enumerate}
    \item ~\eqref{soscond11} implies~\eqref{cd1_1}: 
    if~\eqref{soscond11} are SOS polynomials for all $q\in Q_{\text{init}}$, then for all $(x,\hat{x})$ such that $g_0(q)(x,\hat{x})\geq 0$, one has $\lambda_0^\top(q)g_0(q)\geq 0$, indicating that $-\mathcal{B}_{\Delta(q)}(x,\hat{x})\geq 0$ hold for any $(x,\hat{x},q)\in\tilde{X}_0$, which implies that~\eqref{cd1_1} holds; 
    \item ~\eqref{soscond12} implies~\eqref{cd1_2}: if ~\eqref{soscond11} is SOS polynomials, then for all $(x,\hat{x})$ such that $g_1(x,\hat{x})\geq 0$, one has $\lambda_1^\top(q)g_1\geq 0$. 
    Therefore, one can verify that $\mathcal{B}_{\Delta(\bar{F})}(x,\hat{x})\geq \epsilon >0$ holds for all $(x,\hat{x},q)\in \mathcal{R}\times \bar{F}$ so that~\eqref{cd1_2} holds;  
    \item ~\eqref{soscond13} implies~\eqref{cd1_3}:
    Suppose that~\eqref{soscond13} are SOS polynomials for all $q\in \bar{Q}\setminus \{q_{\text{trap}}$ and $\bar{F}\},q'\in \bar{Q}$.
    Then, for all $(x,\hat{x},u,\hat{u})$ such that $ g_2(q,q')(x,\hat{x},u,\hat{u})\geq 0$ holds, and for all $(u,\hat{u})$ such that $g_u(u,\hat{u})\geq 0$ holds, one gets $\lambda_u^\top\!(q)g_u\geq 0$ and $\lambda_2^\top\!(q) g_2(q,q')\geq 0$. 
    Therefore, one can readily verify that for all $(x,\hat{x},q)\in \mathcal{R}\times \bar{Q}\setminus \{q_{\text{trap},\bar{F}}\}$, $-\mathcal{B}_{\Delta(q')}({f}(x,u),f(\hat{x},\hat{u}))+\mathcal{B}_{\Delta(q)}(x,\hat{x})\geq 0$ so that \eqref{cd1_3} holds.$\hfill\blacksquare$
\end{enumerate}

{\bf Proof for Lemma~\ref{lemma2}}
First, we show that~\eqref{SOS21} are SOS polynomials implies that~\eqref{cd2_1} holds. 
If~\eqref{SOS21} are SOS polynomials for all $q \in Q_{\text{init}}$, then for all $(x,\hat{x})$ such that $g_0(q)(x,\hat{x})\geq 0$, one has $\beta_0^\top(q)g_0(q)\geq 0$ so that $\mathcal{V}_{\Delta(q)}(x,\hat{x})\leq 0$. 
This means that~\eqref{cd2_1} holds. 

Next, we show that~\eqref{SOS22} are SOS polynomials implies that~\eqref{cd2_2} holds.
Suppose that~\eqref{SOS22} are SOS polynomials for any $q\in \bar{Q}\setminus \bar{F}$.
Consider any $(x,\hat{x})$ such that $g_3(q)(x,\hat{x})\geq 0$, one gets $\beta_1^\top(q)g_3\geq 0$ and hence $\mathcal{V}(x,\hat{x},q)\geq \varepsilon_1 >0$.
Therefore, one can conclude that~\eqref{cd2_2} holds.

Finally, we show that~\eqref{SOS23}-\eqref{SOS25} are SOS polynomials implies that~\eqref{cd2_3} holds.
Concretely,~\eqref{SOS24} and~\eqref{SOS25} indicates, respectively, that $\rho^{(q,q')}(x,\hat{x})\in U$, and $\hat{\rho}^{(q,q')}(x,\hat{x})\in U$ hold for any $(x,\hat{x})$ such that $\bar{g}_1(x,\hat{x})\geq 0$, i.e. for all $(x,\hat{x})\in \overline{R}$.
Additionally, if~\eqref{SOS23} are SOS polynomials for all $q\in \bar{Q}\setminus \bar{F}$ and $\forall q' \in \text{Nxt}(q)$, then, for any $(x,\hat{x})$ such that $\bar{g}_1(x,\hat{x})\geq 0$ (therefore $\beta_2^\top(q) \bar{g}_1$), $u_i=\rho^{(q,q')}_i(x,\hat{x})$, and $\hat{u}_i = \hat{\rho}^{(q,q')}_i(x,\hat{x})$, one can readily verify that $-\mathcal{V}_{\Delta(q')}(f(x,u),f(\hat{x},\hat{u}))+\mathcal{V}_{\Delta_q}(x,\hat{x})\geq \varepsilon_2 > 0$.
Hence, one can see that~\eqref{cd2_3} holds, which completes the proof. 
$\hfill\blacksquare$

{\bf Proof for Theorem~\ref{exist_diagnoser}}
We first show the statement regarding \textit{only if} by contraction.
Suppose there exists a ($\delta$,$K$)-diagnoser $D$ satisfying condition (C1) and (C2) as in Definition~\ref{def:diagnoser}, but the dt-CS of interest is not $\delta$-approximate $K$-diagnosable.
On the one hand, the lack of diagnosability indicates that there exists a state run $ \mathbf{x}_{x_0,v}:= (x_0,\ldots,x_k,\ldots)$ with 1) $x_0\in X_0$, 2) $x_k \in X_F$ and $x_i\notin X_F$, $\forall i \in [0,k-1]$, such that there exists another state run $ \hat{\mathbf{x}}_{\hat{x}_0,\hat{v}}:= (\hat{x}_0,\ldots,\hat{x}_k,\ldots)$ for which 1) $\hat{x}_0\in X_0$, 2) $\hat{x}_i\notin X_F$, $\forall i\in[0,k+K]$, 3) $\max_{i\in [0,k+K]} \vert\vert  h(x_i)- h(\hat{x}_i)\vert \vert \leq \delta$. 
On the other hand, in order to satisfy condition (C2) in Definition~\ref{def:diagnoser}, given any imprecise observation $\mathbf{y}^{\delta}_{x_0,\nu}$ within the time horizon $[0,H]$, if there exists a state run $\mathbf{x}'_{x'_0,\nu'}$ such that $\mathbf{x}'_{x'_0,\nu'}(i)\notin X_F$, $\forall i\in[0,H]$, 
and $||\mathbf{y}^{\delta}_{x_0,\nu}(t)- h(\mathbf{x}'_{x_0,\nu}(k))||\leq \delta$, $\forall t\in[0,H]$, one should have $D(\mathbf{y}^{\delta}_{x_0,\nu})=0$.
Then, considering the output run $\mathbf{y}_{x_0,\nu}$ associated with $\mathbf{x}_{x_0,\nu}$ over time horizon $[0,k+K]$, one has $D(\mathbf{y}_{x_0,\nu})=0$ due to the existence of the state run $\hat{\mathbf{x}}_{\hat{x}_0,\hat{v}}$. 
This results in a violation of condition (C1) since $\mathbf{x}_{x_0,\nu}(k)\in \bar{F}$, which is contradictory to the fact that the diagnoser $D$ satisfies conditions (C1) and (C2).
Therefore, the statement regarding \textit{only if} holds.

Next, we proceed with showing the statement regarding \textit{if}.
Suppose the dt-CS $\Sigma$ is $\delta$-approximate $K$-diagnosable.
We first define a set
\begin{align*}
    &\mathcal{Y}(H,\delta) = \big\{\mathbf{y}_{x_0,\nu}\in Y^{H}: \exists \mathbf{x}'_{x_0,\nu} \text{ s.t. } \mathbf{x}'_{x_0,\nu} \notin X_F,||\mathbf{y}^{\delta}_{x_0,\nu}(t)-h(\mathbf{x}'_{x_0,\nu}(t)) ||\leq \delta, \forall t\in[0,H-1] \big\},
\end{align*}
which contains all the output runs that are indistinguishable from some other output runs for which the associated state runs have not entered the fault state set $X_F$.
Based on this set, we construct a diagnoser $D$ such that, given an imprecise observation $\mathbf{y}^{\delta}_{x_0,\nu}$ over time horizon $[0,H]$, 
\begin{align}
    D(\mathbf{y}_{x_0,\nu})= \left\{
    \begin{aligned}
        0, & \quad \quad \quad\text{ if } \mathbf{y}_{x_0,\nu} \in \mathcal{Y}(H,\delta) \\
        1, & \quad \quad \quad\text{ otherwise }
    \end{aligned}
    \right.\label{des_diag}
\end{align}
Then, we prove the statement regarding \textit{if} by showing that such a diagnoser satisfies conditions (C1) and (C2) as in Definition~\ref{def:diagnoser}.
\begin{itemize}
    \item Consider an state run $\mathbf{x}_{x_0,\nu}$ over time horizon $[0,H]$, for any $H\in\mathbb{N}$, such that $\mathbf{x}_{x_0,\nu}(t)\notin X_F$, $\forall t\in[0,H]$, one has $\mathbf{y}^{\delta}_{x_0,\nu}\in \mathcal{Y}(H,\delta)$, with $\mathbf{y}^{\delta}_{x_0,\nu}$ being any imprecise observation as in~\eqref{imprecise} with respect to the observation precision $\delta$. 
    Therefore, one has $D(\mathbf{y}^{\delta}_{x_0,\nu})=0$, indicating that condition (C2) in Definition~\ref{def:diagnoser} is satisfied.
    \item Consider an state run $\mathbf{x}_{x_0,\nu}$ over time horizon $[0,k+K]$, such that $\mathbf{x}_{x_0,\nu}(k)\in X_F$ and $\mathbf{x}_{x_0,\nu}(t)\notin X_F$ for all $t\in[0,k-1]$.
    Since the dt-CS $\Sigma$ is $\delta$-approximate $K$-diagnosable, for all $\mathbf{x}'_{x_0,\nu}$ such that $\mathbf{x}'_{x_0,\nu}(t)\notin X_F$, $\forall t\in [0,k+K]$, there exists $t''\in [0,k+K]$, $||h(\mathbf{x}_{x_0,\nu}(t''))-h(\mathbf{x}'_{x_0,\nu}(t'')) ||>0$.
    This immediately indicates that for any imprecise observation $\mathbf{y}^{\delta}_{x_0,\nu}$ associated with $\mathbf{x}_{x_0,\delta}$, one has $\mathbf{y}^{\delta}_{x_0,\nu}\notin \mathcal{Y}(H,\delta)$.
    Therefore, one has $D(\mathbf{y}^{\delta}_{x_0,\nu})=1$ according to~\eqref{des_diag}, indicating that condition (C1) in Definition~\ref{def:diagnoser} is satisfied, which completes the proof.$\hfill\blacksquare$
\end{itemize}

\begin{figure*}
\rule[0pt]{\textwidth}{0.05em}
\begin{small}
\begin{align}
	&\mathcal{B}_{SMT}(x,\hat{x}, q):=\nonumber\\
	&\left\{\begin{aligned}
&-0.1459 x_1 ^ 2 + 0.0005 x_1   x_2 + 0.0002 x_1   \hat{x}_1 + 0.0005 x_1   \hat{x}_2 + 7.2650 x_1 + 0.0942 x_2 ^ 2 + 0.0001 x_2   \hat{x}_1 - 0.1894 x_2   \hat{x}_2  \\
&\quad \quad \quad \quad \quad \quad  \quad \quad \quad  + 0.0100 x_2 - 0.0008 \hat{x}_1 ^ 2 + 0.0001 \hat{x}_1   \hat{x}_2 + 0.0303 \hat{x}_1+ 0.0942 \hat{x}_2 ^ 2 + 0.0100 \hat{x}_2 - 90.1333, \text{ if } q = {q}_0;\\
&0.1049 x_1 ^ 2 - 0.0034 x_1   x_2 - 0.0009 x_1   \hat{x}_1 - 0.0034 x_1   \hat{x}_2 - 5.0597 x_1 + 0.3635 x_2 ^ 2 - 0.0003 x_2   \hat{x}_1 - 0.7281 x_2   \hat{x}_2 \\
&\quad \quad \quad \quad \quad \quad  \quad \quad \quad   + 0.1188 x_2 - 0.0007 \hat{x}_1 ^ 2 -0.0003 \hat{x}_1   \hat{x}_2 + 0.0702 \hat{x}_1 + 0.3635 \hat{x}_2 ^ 2 + 0.1188 \hat{x}_2 + 59.5153, \text{ if } q = {q}_1;\\
&0.1012 x_1 ^ 2 - 0.0041 x_1   x_2 - 0.0021 x_1   \hat{x}_1 - 0.0041 x_1   \hat{x}_2 - 4.8182 x_1 + 0.3467 x_2 ^ 2 - 0.0005 x_2   \hat{x}_1 - 0.6949 x_2   \hat{x}_2 \\
&\quad \quad \quad \quad \quad \quad  \quad \quad \quad   + 0.1504 x_2 - 0.0011 \hat{x}_1 ^ 2  -0.0005 \hat{x}_1   \hat{x}_2 + 0.1273 \hat{x}_1 + 0.3467 \hat{x}_2 ^ 2 + 0.1504 \hat{x}_2 + 55.0563, \text{ if } q = {q}_2;\\
&0.0883 x_1 ^ 2 - 0.0042 x_1   x_2 - 0.0027 x_1   \hat{x}_1 - 0.0042 x_1   \hat{x}_2 - 4.1521 x_1 + 0.3107 x_2 ^ 2 - 0.0006 x_2   \hat{x}_1 - 0.6234 x_2   \hat{x}_2 \\
&\quad \quad \quad \quad \quad \quad  \quad \quad \quad    + 0.1643 x_2 - 0.0013 \hat{x}_1 ^ 2 -0.0006 \hat{x}_1   \hat{x}_2 + 0.1614 \hat{x}_1 + 0.3107 \hat{x}_2 ^ 2 + 0.1643 \hat{x}_2 + 45.9998, \text{ if } q = {q}_3;\\
&0.0775 x_1 ^ 2 - 0.0037 x_1   x_2 - 0.0036 x_1   \hat{x}_1 - 0.0037 x_1   \hat{x}_2 - 3.6148 x_1 + 0.2621 x_2 ^ 2 - 0.0007 x_2   \hat{x}_1 - 0.5263 x_2   \hat{x}_2 \\
&\quad \quad \quad \quad \quad \quad  \quad \quad \quad  + 0.1592 x_2 - 0.0015 \hat{x}_1 ^ 2 -0.0007 \hat{x}_1   \hat{x}_2 + 0.1973 \hat{x}_1 + 0.2621 \hat{x}_2 ^ 2 + 0.1592 \hat{x}_2 + 38.9878, \text{ if } q = {q}_4;\\
&0.0531 x_1 ^ 2 - 0.0035 x_1   x_2 - 0.0035 x_1   \hat{x}_1 - 0.0035 x_1   \hat{x}_2 - 2.4083 x_1 + 0.1793 x_2 ^ 2 - 0.0007 x_2   \hat{x}_1 - 0.3606 x_2   \hat{x}_2 \\
&\quad \quad \quad \quad \quad \quad  \quad \quad \quad   + 0.1538 x_2 - 0.0016 \hat{x}_1 ^ 2 -0.0007 \hat{x}_1   \hat{x}_2 + 0.1962 \hat{x}_1 + 0.1793 \hat{x}_2 ^ 2 + 0.1538 \hat{x}_2 + 24.0745, \text{ if } q = {q}_5;\\
&-0.0028 x_1 ^ 2 + 0.0029 x_1   x_2 + 0.0058 x_1   \hat{x}_1 + 0.0029 x_1   \hat{x}_2 + 0.1072 x_1 + 0.0033 x_2 ^ 2 + 0.0006 x_2   \hat{x}_1 - 0.0059 x_2   \hat{x}_2 \\
&\quad \quad \quad \quad \quad \quad  \quad \quad \quad    - 0.0912 x_2 - 0.0033 \hat{x}_1 ^ 2 + 0.0006 \hat{x}_1   \hat{x}_2 + 0.0171 \hat{x}_1 + 0.0033 \hat{x}_2 ^ 2 - 0.0912 \hat{x}_2 - 4.1225, \text{ if } q = {q}_{\text{trap}};\\
&0.0284 x_1 ^ 2 - 0.0025 x_1   x_2 - 0.0025 x_1   \hat{x}_1 - 0.0025 x_1   \hat{x}_2 - 1.2449 x_1 + 0.1155 x_2 ^ 2 - 0.0006 x_2   \hat{x}_1 - 0.2326 x_2   \hat{x}_2 \\
&\quad \quad \quad \quad \quad \quad  \quad \quad \quad    + 0.1145 x_2 - 0.0013 \hat{x}_1 ^ 2 -0.0006 \hat{x}_1   \hat{x}_2 + 0.1534 \hat{x}_1 + 0.1155 \hat{x}_2 ^ 2 + 0.1145 \hat{x}_2 + 11.0148, \text{ if } q = \bar{F};\\
	\end{aligned}\right.\label{Bcase}
\end{align}
\end{small}
\rule[0pt]{\textwidth}{0.05em}
\end{figure*}

\begin{figure*}
	
	\begin{small}
		\begin{align}
			&\mathcal{B}_{SOS}(x,\hat{x}, q):=\nonumber\\
			&\left\{\begin{aligned}
				& 61.9793 + 52.4632 x_1 + 30.0867 x_2  + 45.4627 \hat{x}_1  + 36.2866 \hat{x}_2 + 
				75.4901 x_1^2 - 104.6505 x_1 x_2  + 76.7151 x_2^2 - 112.4098 x_1 \hat{x}_1 \\
				& \quad \quad \quad \quad \quad \quad  - 101.8547 x_2 \hat{x}_1 +
				75.9589 \hat{x}_1^2 -100.8383 x_1 \hat{x}_2  - 115.4499 x_2 \hat{x}_2  - 104.7739 \hat{x}_1 \hat{x}_2 + 74.8508 \hat{x}_2^2, \text{ if } q = {q}_0;\\
				& 24.1637 + 2.0924 x_1 + 8.9078 x_2  + 20.9364 \hat{x}_1  + 16.4481 \hat{x}_2 + 90.5688 x_1^2 - 113.9268 x_1 x_2  + 80.0829 x_2^2  - 107.6873 x_1 \hat{x}_1 \\ 
				& \quad \quad \quad \quad \quad \quad  - 105.7055 x_2 \hat{x}_1 + 84.5255 \hat{x}_1^2 -107.0687 x_1 \hat{x}_2  - 116.2163 x_2 \hat{x}_2  - 106.6658 \hat{x}_1 \hat{x}_2 + 76.3952 \hat{x}_2^2, \text{ if } q = {q}_1;\\
				& 5.6908 + 13.8516 x_1 + 13.4998 x_2  + 25.8271 \hat{x}_1  + 12.1339 \hat{x}_2 + 10.4628 x_1^2 - 102.3958 x_1 x_2  + 6.8278 x_2^2  - 97.1537 x_1 \hat{x}_1 \\ 
				& \quad \quad \quad \quad \quad \quad  - 98.7575 x_2 \hat{x}_1 + 5.0966 \hat{x}_1^2 -99.7841 x_1 \hat{x}_2  - 102.4821 x_2 \hat{x}_2  - 100.9902 \hat{x}_1 \hat{x}_2 + 5.1242 \hat{x}_2^2, \text{ if } q = {q}_2;\\
				& 3.4248 + 10.0281 x_1 + 9.6047 x_2  + 24.1036 \hat{x}_1  + 10.4827 \hat{x}_2 -17.767 x_1^2 - 129.2525 x_1 x_2  + 19.9533 x_2^2  - 128.2189 x_1 \hat{x}_1 \\ 
				& \quad \quad \quad \quad \quad \quad  - 125.6612 x_2 \hat{x}_1 -24.1346 \hat{x}_1^2 -126.3442 x_1 \hat{x}_2  - 130.766 x_2 \hat{x}_2  - 127.7656 \hat{x}_1 \hat{x}_2 + 22.5702 \hat{x}_2^2, \text{ if } q = {q}_3;\\
				& 0.084 + 4.9894 x_1 + 4.636 x_2  + 20.4027 \hat{x}_1  + 7.67 \hat{x}_2 -51.3395 x_1^2 - 165.9146 x_1 x_2  + 52.2193 x_2^2  - 168.9452 x_1 \hat{x}_1 \\ 
				& \quad \quad \quad  \quad \quad  - 162.4919 x_2 \hat{x}_1 -58.7638 \hat{x}_1^2 -162.2771 x_1 \hat{x}_2  - 168.6144 x_2 \hat{x}_2  - 164.0869 \hat{x}_1 \hat{x}_2 + 55.1961 \hat{x}_2^2, \text{ if } q = {q}_4;\\
				& 14.5433 + 0.3807 x_1 + 0.4084 x_2  + 14.7334 \hat{x}_1  + 3.9185 \hat{x}_2 -107.1316 x_1^2 - 232.5885 x_1 x_2  + 106.7981 x_2^2  - 233.5661 x_1 \hat{x}_1 \\ 
				& \quad \quad \quad \quad \quad   - 229.9268 x_2 \hat{x}_1 -115.5868 \hat{x}_1^2 -229.3328 x_1 \hat{x}_2  - 232.0117 x_2 \hat{x}_2  - 231.8204 \hat{x}_1 \hat{x}_2 + 110.4617 \hat{x}_2^2, \text{ if } q = {q}_5;\\
				& -6.2326 + 12.3431 x_1 + 12.6831 x_2  + 12.4161 \hat{x}_1  + 11.9542 \hat{x}_2 + 0.5714 x_1^2 - 0.9295 x_1 x_2  + 0.5441 x_2^2  - 0.7124 x_1 \hat{x}_1 \\ 
				& \quad \quad \quad \quad \quad \quad  - 0.6987 x_2 \hat{x}_1 + 0.5522 \hat{x}_1^2 -0.7316 x_1 \hat{x}_2  - 0.7252 x_2 \hat{x}_2  - 0.9339 \hat{x}_1 \hat{x}_2 + 0.6022 \hat{x}_2^2, \text{ if } q = {q}_{\text{trap}};\\
				&-10.5989 + 28.1246 x_1 + 28.5379 x_2  + 31.5337 \hat{x}_1  + 30.2472 \hat{x}_2 -197.7538 x_1^2 - 386.1975 x_1 x_2  + 201.7908 x_2^2   \\ 
				& - 377.34 x_1 \hat{x}_1 - 389.3004 x_2 \hat{x}_1 -205.1699 \hat{x}_1^2 -391.8408 x_1 \hat{x}_2  - 386.5924 x_2 \hat{x}_2  - 402.3205 \hat{x}_1 \hat{x}_2 + 212.0586 \hat{x}_2^2, \text{ if } q = \bar{F};\\
			\end{aligned}\right.\label{Bcase-2}
		\end{align}
	\end{small}
	\rule[0pt]{\textwidth}{0.05em}
\end{figure*}

\begin{figure*}
	\begin{small}
		\begin{align}
			&\mathcal{V}_{SMT}(x,\hat{x}, q):=\nonumber\\
			&\left\{\begin{aligned}
				& 0.0008 x_1^2  - 0.5816 x_1 x_2 - 0.0009 x_1 \hat{x}_1 + 0.581 x_1 \hat{x}_2  - 0.002 x_1 +   25.8503 x_2^2 + 0.6081 x_2 \hat{x}_1 - 51.7281 x_2 \hat{x}_2  - 0.1994 x_2 \\
                & \quad \quad \quad \quad \quad \quad  - 0.0001 \hat{x}_1^2 - 0.6081 \hat{x}_1 \hat{x}_2 +0.0272 \hat{x}_1  + 25.8777 \hat{x}_2^2 + 0.2156 \hat{x}_2 - 0.4762  , \text{ if } q = {q}_0;\\
				& 0.0409 x_1^2  - 0.0564 x_1 x_2 - 0.026 x_1 \hat{x}_1 + 0.0555 x_1 \hat{x}_2  - 2.6383 x_1 +   2.6015 x_2^2 + 0.0705 x_2 \hat{x}_1 - 5.2038 x_2 \hat{x}_2  - 4.4242 x_2 \\
                & \quad \quad \quad \quad \quad \quad  - 0.0286 \hat{x}_1^2 - 0.0706 \hat{x}_1 \hat{x}_2 +0.7165 \hat{x}_1  + 2.6016 \hat{x}_2^2 + 4.4325 \hat{x}_2 - 23.7729, \text{ if } q = {q}_1;\\
				& 0.0533 x_1^2  - 0.069 x_1 x_2 - 0.0372 x_1 \hat{x}_1 + 0.0695 x_1 \hat{x}_2  - 3.5831 x_1 +   2.1818 x_2^2 + 0.1202 x_2 \hat{x}_1 - 4.3665 x_2 \hat{x}_2  - 2.5315 x_2 \\
                & \quad \quad \quad \quad \quad \quad  - 0.0395 \hat{x}_1^2 - 0.1206 \hat{x}_1 \hat{x}_2 +0.9576 \hat{x}_1  + 2.1837 \hat{x}_2^2 + 2.5801 \hat{x}_2 - 33.3048, \text{ if } q = {q}_2;\\
				& 0.0495 x_1^2  - 0.0267 x_1 x_2 - 0.151 x_1 \hat{x}_1 + 0.0264 x_1 \hat{x}_2  - 6.1248 x_1 +   3.0936 x_2^2 + 0.3715 x_2 \hat{x}_1 - 6.1864 x_2 \hat{x}_2  - 4.8224 x_2 \\
                & \quad \quad \quad \quad \quad \quad  - 0.1299 \hat{x}_1^2 - 0.372 \hat{x}_1 \hat{x}_2 +2.3982 \hat{x}_1  + 3.0924 \hat{x}_2^2 + 4.8275 \hat{x}_2 - 48.9751, \text{ if } q = {q}_3; \\
				& 0.0002 x_1^2  - 0.0001 x_1 x_2 - 0.0001 x_1 \hat{x}_1 + 0.0001 x_1 \hat{x}_2  - 0.0086 x_1  - 0.0002 x_2 \hat{x}_2  - 0.0004 x_2 \\
                & \quad \quad \quad \quad \quad \quad \quad \quad \quad \quad  - 0.0001 \hat{x}_1 \hat{x}_2 +0.0019 \hat{x}_1  + 0.0001 \hat{x}_2^2 + 0.0054 \hat{x}_2 - 0.1784, \text{ if } q = {q}_{\text{trap}};\\
				& -0.0143 x_1^2  - 0.0074 x_1 x_2 - 0.0066 x_1 \hat{x}_1 + 0.0073 x_1 \hat{x}_2  - 0.0434 x_1 +   0.0007 x_2^2 + 0.0009 x_2 \hat{x}_1 - 0.0006 x_2 \hat{x}_2  - 0.1176 x_2 \\
                & \quad \quad \quad \quad \quad \quad  - 0.0009 \hat{x}_1^2 - 0.0009 \hat{x}_1 \hat{x}_2 +0.1176 \hat{x}_1  + 0.0006 \hat{x}_2^2 + 0.1195 \hat{x}_2 - 3.886, \text{ if } q = \bar{F};\\
			\end{aligned}\right.\label{Bcase-V-1}
		\end{align}
	\end{small}
	\rule[0pt]{\textwidth}{0.05em}
\end{figure*}

\begin{figure*}
	
	\begin{small}
		\begin{align}
			&\mathcal{V}_{SOS}(x,\hat{x}, q):=\nonumber\\
			&\left\{\begin{aligned}
				& 90.873 + 128.5902 x_1 + 126.1268 x_2  + 123.4983 \hat{x}_1  + 118.9705 \hat{x}_2 + 175.8664 x_1^2 + 2.1959 x_1 x_2  + 170.959 x_2^2   \\
				& \quad- 362.3845 x_1 \hat{x}_1 - 10.1361 x_2 \hat{x}_1 + 191.9901 \hat{x}_1^2 -5.235 x_1 \hat{x}_2  - 346.6525 x_2 \hat{x}_2  - 30.6354 \hat{x}_1 \hat{x}_2 + 180.5615 \hat{x}_2^2, \text{ if } q = {q}_0;\\
				& 21.0653 + 144.2286 x_1 + 139.518 x_2  + 147.671 \hat{x}_1  + 148.1492 \hat{x}_2 + 132.3899 x_1^2 + 72.9988 x_1 x_2  + 130.6841 x_2^2  \\
				&  - 172.7302 x_1 \hat{x}_1 - 44.1475 x_2 \hat{x}_1 + 118.5449 \hat{x}_1^2 -38.0244 x_1 \hat{x}_2  - 166.7411 x_2 \hat{x}_2  - 31.0273 \hat{x}_1 \hat{x}_2 + 110.8106 \hat{x}_2^2, \text{ if } q = {q}_1;\\
				& 60.0098 + 116.1265 x_1 + 116.4412 x_2  + 116.183 \hat{x}_1  + 117.3591 \hat{x}_2 + 121.4941 x_1^2 + 65.6934 x_1 x_2  + 120.262 x_2^2   \\
				& \quad- 150.5795 x_1 \hat{x}_1 - 45.7302 x_2 \hat{x}_1 + 114.0595 \hat{x}_1^2 -39.9788 x_1 \hat{x}_2  - 145.5936 x_2 \hat{x}_2  - 40.8048 \hat{x}_1 \hat{x}_2 + 107.24 \hat{x}_2^2, \text{ if } q = {q}_2;\\
				& -19.3845 + 118.5786 x_1 + 117.8461 x_2  + 118.0923 \hat{x}_1  + 116.5122 \hat{x}_2 + 58.8333 x_1^2 + 30.7074 x_1 x_2  + 57.7209 x_2^2   \\
				& \quad\quad\,- 84.1498 x_1 \hat{x}_1 - 15.108 x_2 \hat{x}_1 + 56.0272 \hat{x}_1^2 -15.7012 x_1 \hat{x}_2  - 84.5201 x_2 \hat{x}_2  - 23.8536 \hat{x}_1 \hat{x}_2 + 55.2207 \hat{x}_2^2, \text{ if } q = {q}_3; \\
				& -83.6295 + 146.2733 x_1 + 146.1867 x_2  + 146.8514 \hat{x}_1  + 144.0165 \hat{x}_2 -1.507 x_1^2 + 7.1393 x_1 x_2  + 1.5837 x_2^2  - 31.5865 x_1 \hat{x}_1 \\
				& \quad \quad \quad \quad \quad \quad \quad\quad\, - 24.3528 x_2 \hat{x}_1 -1.8947 \hat{x}_1^2 -25.1087 x_1 \hat{x}_2  - 33.2263 x_2 \hat{x}_2  - 7.0456 \hat{x}_1 \hat{x}_2 + 4.0396 \hat{x}_2^2, \text{ if } q = {q}_{\text{trap}};\\
				& 56.6331 + 137.5805 x_1 + 137.7372 x_2  + 137.4526 \hat{x}_1  + 137.6363 \hat{x}_2 -22.7515 x_1^2 + 3.2284 x_1 x_2  + 24.1708 x_2^2  - 2.7251 x_1 \hat{x}_1 \\
				& \quad \quad \quad \quad \quad \quad \quad\quad\quad\quad \quad- 2.2209 x_2 \hat{x}_1 -22.7993 \hat{x}_1^2 -1.6331 x_1 \hat{x}_2  - 7.2252 x_2 \hat{x}_2  - 3.6501 \hat{x}_1 \hat{x}_2 + 25.489 \hat{x}_2^2, \text{ if } q = \bar{F};\\
			\end{aligned}\right.\label{Bcase-V-2}
		\end{align}
	\end{small}
	\rule[0pt]{\textwidth}{0.05em}
\end{figure*}

{\bf Proof for Theorem~\ref{observer}}
We first show \text{(i)} by contradiction.
Suppose that one has $\mathbf{x}_{x_0,\nu}(k')\in X_F$ , but $\forall i\in [0,k'+K]$, $M(i)\neq \emptyset$.
Then, there exists a state run $\mathbf{x}_{\bar{x}_0,\bar{v}}$, such that $\mathbf{x}_{\bar{x}_0,\bar{v}}(j)\in M(j)$ for all $j\in[0,k'+K]$.
In other words, one has $\mathbf{x}_{\bar{x}_0,\bar{v}}(j)\notin X_F$ for all $j\in[0,k'+K]$, while one has
\begin{align}
	\max_{j\in [0,k+K]} \vert\vert  h(\mathbf{x}_{\bar{x}_0,\bar{v}}(j))- h(\mathbf{x}_{x_0,\nu}(j))\vert \vert \leq \delta.\label{eq5.2.1}
\end{align}
Note that~\eqref{eq5.2.1} results in a contradiction to the fact that the dt-CS $\Sigma$ is $\delta$-approximate $K$-diagnosable considering Definition~\ref{def:diagnosability}.

Now, we proceed with showing \text{(ii)}.
First, suppose that $\nexists k'\in [0,\mathsf{i}]$ such that $\mathbf{x}_{x_0,\nu}(k')\in X_F$.
Then, one can readily verify that $\forall k'\in [0,\mathsf{i}]$, $\mathbf{x}_{x_0,\nu}(k')\in M(k')$.
In other words, one can conclude that $M(k')\neq \emptyset$ for all $k'\in [0,\mathsf{i}]$ since $\forall k'\in [0,\mathsf{i}]$, the set $M(k')$ at least contains $\mathbf{x}_{x_0,\nu}(k')$.
This results in a contradiction to the fact that $M(\mathsf{i})=\emptyset$.
Therefore, one can conclude that 
\begin{align}
\exists k'\in [0,\mathsf{i}], \mathbf{x}_{x_0,\nu}(k')\in X_F. \label{step1}
\end{align}
Based on~\eqref{step1}, we complete the proof for \textit{(ii)} via showing that $\nexists \mathsf{k}\in [0,\mathsf{i}-K-1]$ such that $\mathbf{x}_{x_0,\nu}(k')\in X_F$.
Suppose that $\exists \mathsf{k}\in [0,\mathsf{i}-K-1]$ such that $\mathbf{x}_{x_0,\nu}(\mathsf{k})\in X_F$.
By applying \textit{(ii)}, there should exist $\mathsf{j}\in[0,\mathsf{k}+K]$ such that $M(\mathsf{j})=\emptyset$. 
This is contradictory to the fact that $M(j)\neq \emptyset$, $\forall j\in[0,\mathsf{i}-1]$ since $\mathsf{k}+K< \mathsf{i}$.
Therefore, one can conclude that 
\begin{align}
\nexists \mathsf{k}\in [0,\mathsf{i}-K-1], \mathbf{x}_{x_0,\nu}(\mathsf{k})\in X_F. \label{step2}
\end{align}
Then, the proof is completed considering~\eqref{step1} and~\eqref{step2}.
$\hfill\blacksquare$

\end{document}